\documentclass[11pt]{article}
\usepackage[numbers, square]{natbib}
\usepackage{graphicx}

\usepackage[margin=10pt,font=small,labelfont=bf,labelsep=period]{caption}

\usepackage{latexsym}

\def\rm{\textrm}
\def\bf{\textbf}

\def\bm{$\textrm d_{bm\,}$}
\def\be{\begin{equation}}
\def\ee{\end{equation}}
\def\bea{\begin{eqnarray}}
\def\eea{\end{eqnarray}}

\def\case#1/#2{\textstyle\frac{#1}{#2}}

\begin{document}

\vspace{.7in}

\begin{center}
\Large\bf {SHAPE DYNAMICS. An Introduction}\footnote{To appear in the refereed
proceedings of the conference Quantum Field Theory and Gravity (Regensburg,
2010).}

\vspace{.2in} \normalsize \large{\textbf{Julian
Barbour}}\,\footnote{Email: Julian.Barbour@physics.ox.ac.uk}

\end{center}

\normalsize

\vspace{.2in}

{\textbf{Abstract.} Shape dynamics is a completely background-independent
universal framework of dynamical theories from which all absolute
elements have been eliminated. For particles, only the variables that describe
the shapes of the instantaneous particle configurations are dynamical. In the
case of Riemannian three-geometries, the only dynamical variables are the parts
of the metric that determine angles. The local scale factor plays no role.
This leads to a shape-dynamic theory of gravity in which the four-dimensional
diffeomorphism invariance of general relativity is replaced by
three-dimensional diffeomorphism invariance and three-dimensional conformal
invariance. Despite this difference of symmetry groups, it is remarkable that
the predictions of the two theories -- shape dynamics and general relativity --
agree on spacetime foliations by hypersurfaces of constant mean extrinsic
curvature. However, the two theories are distinct, with shape dynamics
having a much more restrictive set of solutions. There are indications that the
symmetry group of shape dynamics makes it more amenable to quantization and
thus to the creation of quantum gravity.
This introduction presents in simple terms the arguments for shape dynamics,
its implementation techniques, and a survey of existing results.

\tableofcontents

\section{Introduction}

One of Einstein's main aims in creating general relativity was to implement
Mach's idea
\cite{Mach1883, Mach1960} that dynamics should use only relative quantities and
that inertial motion as expressed in Newton's first law should arise, not as
an effect of a background absolute space, but from the dynamical effect of the
universe as a
whole. Einstein called this \emph{Mach's principle}
\cite{Einstein1918}. However, as he explained later \cite{Einstein1918a,
Barbour1995} (p.\,186), he found it impractical to realize Mach's principle
directly and was forced to use coordinate systems. This has obscured the extent
to which and how general relativity is a background-independent theory. My aim
in this paper is to present a universal framework for the \emph{direct} 
and \emph{explicit} creation of completely background-independent theories. 

I shall show that this leads to a theory of gravity, \emph{shape dynamics}, that
is distinct from general
relativity because it is based on a different symmetry group, according to which
only the local shapes of Riemannian 3-geometries are dynamical.
Nevertheless, it is remarkable that the two theories have a nontrivial
`intersection', agreeing exactly in spatially closed universes whenever and
wherever Einsteinian spacetimes admit foliation by hypersurfaces of constant
mean extrinsic curvature. However, many solutions of general relativity that
appear manifestly unphysical, such as those with closed timelike curves, are
not allowed in shape dynamics. In addition, it appears that the structure of
shape dynamics makes it significantly more amenable to quantization than
general relativity.

This is not the only reason why I hope the reader will take an interest in
shape dynamics. The question of whether motion is absolute or relative has a
venerable history \cite{Barbour1989, Barbour2001}, going back to long before
Newton made it famous when he formulated dynamics in terms of absolute space and
time \cite{Newton1962}. What is ultimately at stake is the definition of
position and, above all, velocity. This has abiding relevance in our
restless universe. I shall show that it is possible to eliminate every
vestige of Newtonian absolutes except for just one. But this solitary remnant
is hugely important: it allows the universe to expand. Shape dynamics highlights
this remarkable fact.

This introduction will be to a large degree heuristic and based on Lagrangian
formalism. A more rigorous Hamiltonian formulation of shape dynamics better
suited to calculations and quantum-gravity applications was recently discovered
in \cite{Gomes2011} (a simplified treatment is in \cite{Gomes2011a}. Several
more papers developing the Hamiltonian formulation in directions that appear
promising from the quantum-gravity perspective are in preparation. A dedicated
website (shapedynamics.org) is under construction; further background
information can be found at platonia.com.

The contents list obviates any further introduction, but a word on
terminology will help. Two distinct
meanings of \emph{relative} are often confused. Mach
regarded inter-particle separations as relative quantities; in Einstein's
theories, the division of spacetime into space and time is made relative to an
observer's coordinate system. To avoid confusion, I use \emph{relational} in
lieu of Mach's notion of relative. 

\section{The Relational Critique of Newton's Dynamics}

\subsection{Elimination of redundant structure}

Newton's First Law states: ``Every body continues in its state of rest or
uniform motion in a right line unless it is compelled to change that state by
forces impressed on it.'' Since the (absolute) space in which the body's motion
is said to be straight and the (absolute) time that measures its uniformity are
both invisible, this law as stated is clearly problematic. Newton knew this and
argued in his Scholium in the \emph{Principia} \cite{Newton1962} that his
invisible
absolute motions could be deduced from visible relative motions.
This can be done but requires more relative data than
one would expect if only directly
observable initial data governed the dynamics. As we shall see, this
fact, which is not widely
known, indicates how mechanics can be reformulated with less
kinematic structure than Newton assumed and simultaneously be made more
predictive. It is possible to create a framework that fully resolves the
debate about the nature of motion. In this framework, \emph{the
fewest possible observable
initial data determine the observable evolution}.\footnote{The notion of what
is observable is not
unproblematic. For now it will suffice that inter-particle
separations are more readily
observed than positions in invisible space.}

I show first that all candidate relational configurations\footnote{We shall see
(Sec.~4) that the foundation of dynamics on instantaneous extended
configurations, rather than point events, is perfectly compatible with
Einsteinian relativity. \label{worry}} of the universe have structures
determined by a Lie group, which may be termed their \emph{structure group}. The
existence of such a group is decisive. It leads directly to a natural way to
achieve the aim just formulated and to a characteristic universal structure of
dynamics applicable to which is not
widely known, 
a large class of systems.  It
is present in modern gauge theories and, in its most perfect
form, in general relativity. However, the relational core of these theories is
largely hidden because their
formulation retains redundant kinematic structure.

To identify the mismatch that shape dynamics aims to eliminate, the first step
is to
establish the essential structure that Newtonian dynamics employs. It
will be sufficient to consider $N, N\ge 3,$ point
particles interacting through Newtonian gravity. In an assumed inertial frame
of reference, each particle $a, a=1,...,N,$
has coordinates $x^i_a(t), i=x,y,z,$ that
depend on $t$, the Newtonian time. The
$x^i_a$'s and $t$ are all assumed to be observable. The particles, assumed
individually identifiable, also
have constant masses $m_a$. For the purposes of our discussion,
they can be assumed known. 

Let us now eliminate potentially redundant structure.
Newton granted
that only the inter-particle separations $r_{ab}$, assumed to be `seen' all at
once, are observable. In fact, this presupposes an external (absolute) ruler.
Closer to empirical reality are the dimensionless ratios
\begin{equation}
\tilde r_{ab}:={r_{ab}\over
R_{rmh}},~R_{rmh}:=\sqrt{\sum_{a<b}r_{ab}^2},\label{rmh}
\end{equation}
where $R_{rmh}$ is the root-mean-harmonic separation. It is closely related
to the centre-of-mass moment of inertia $I_{cms}$:
\begin{equation}
I_{cms}:=\sum_am_a\textbf{x}^a\cdot\textbf{x}^a\equiv{1\over
M}\sum_{a<b}m_am_br_{ ab } ^2,~M:=\sum_am_a. \label{cms}
\end{equation}
The system has the `size' $\sqrt {I_{cms}}$ if we grant
a scale,  but we do not and take the instantaneous sets
$\{\tilde r_{ab}\}$ of scale-free ratios $\tilde r_{ab}$ to be
our raw data. They are
`snapshots' of the \emph{instantaneous shapes} of the system. The time
$t$ too is unobservable. There is no clock hung up in space, just the particles
moving relative to each other. All that we have are the sets $\{\tilde
r_{ab}\}$. The totality of such sets is \emph{shape space} $\textrm
Q_{ss}^N$, which only exists for $N\ge 3$.\footnote{A single point is not a
shape, and the distance between
two particles can be scaled to any value, so nothing dimensionless
remains to define a shape. Also the configuration in which all
particles coincide is not a shape and does not belong to shape space.}
The number of dimensions of $\textrm Q_{ss}^N$ is $3N-7$: from the $3N$
Cartesian coordinates, six are subtracted because Euclidean
translations and rotations do not change
the $r_{ab}$'s and the seventh because the $\{\tilde r_{ab}\}$'s are scale
invariant. 

Shape space is our key concept. Mathematically, we reach it
through a succession of spaces, the
first being the $3N$-dimensional \emph{Cartesian configuration space}
$\textrm Q^N$. In it, all configurations that are
carried into each other by translations $\textsf t$ in $\textsf T$, the group of
Euclidean translations, belong to a common \emph{orbit}
of $\textsf T$. Thus, $\textsf T$ decomposes $\rm Q^N$ into
its group orbits, which are defined to be the points
of the $3N-3$-dimensional \emph{quotient space} $\textrm T^N:=\textrm
Q^N/\textsf T$. This first quotienting to $\textrm T^N$ is relatively
trivial. More significant is the further quotienting by the rotation group
$\textsf R$ to the $3N-6$-dimensional \emph{relative configuration
space} $\textrm Q_{rcs}^N:=\textrm Q^N/\textsf T\textsf R$ \cite{Barbour1982}.
The final
quotienting by the dilatation (scaling) group $\textsf S$ leads to shape space
$\textrm Q_{ss}^N:=\textrm Q^N/\textsf{TRS}$ \cite{Barbour2003}. The groups
$\textsf T$ and
$\textsf R$ together form the \emph{Euclidean group}, while the inclusion of
$\textsf S$ yields the \emph{similarity group}. The orbit
of a group is a space with as many dimensions as the number of elements
that specify a group element. The orbits of $\sf S$
thus have seven dimensions (Fig.~1).

\begin{figure}
 \begin{center}
 \includegraphics[width=11cm]{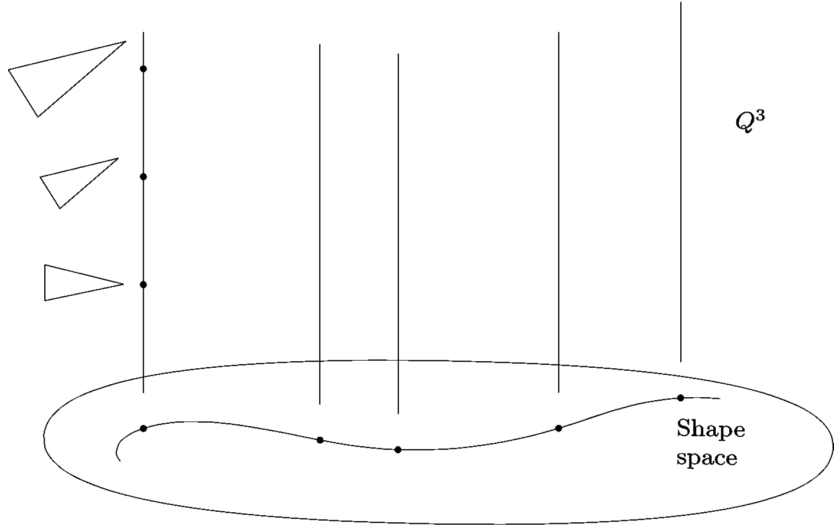}

\caption{Shape space for the 3-body problem is obtained by decomposing the
Newtonian configuration space $Q^3$ into orbits of the similarity group
$\textsf S$. The points on any given vertical line (any group orbit) correspond
to all
possible representations in Euclidean space of one of the possible shapes of the
triangle formed by the three particles. Each such shape is represented below
its orbit as a point in shape space. Each orbit is actually a seven-dimensional
space. The effects of rotation and scaling are shown.}

 \end{center}

\end{figure}

The groups $\textsf{T},\textsf{R},\textsf{S}$ are \emph{groups of
motion}, or \emph{Lie groups} (groups that are
simultaneously manifolds, i.e., their elements are parametrized by
continuous parameters). If we have a configuration $q$ of $N$ particles
in Euclidean space, $q\in Q^N$, we can `move it around' with $\textsf
T$ or $\textsf R$ or `change its size' with $\textsf S$. This
intuition was the
basis of Lie's work. It formalizes the fundamental geometrical
notions of \emph{congruence} and \emph{similarity}. Two figures are
\emph{congruent} if they can be brought to exact overlap by a combination of
translations and rotations and \emph{similar} if dilatations are allowed as
well.

Relational particle dynamics can be formulated in any of the quotient spaces
just considered. Intuition
suggests that the dynamics of an `island universe' in
Euclidean space should deal solely with its possible shapes. The similarity
group is then the fundamental structure group.\footnote{One might want
to go further
and consider the general linear group, under which angles are no longer
invariant. I will consider this possibility later.} This leads to particle shape
dynamics and by analogy to the conformal geometrodynamics that will be
considered in the second part of the paper.

Lie groups and their infinite-dimensional generalizations are
fundamental in modern mathematics and theoretical physics. They play a dual role
in shape dynamics, first in indicating how potentially redundant structure can
be pared away and, second, in
providing the tool to create theories that are
relationally perfect, i.e., free of the mismatch noted above.
Moreover, because Lie groups, as groups that are simultaneously manifolds, have
a common underlying structure and are ubiquitous, they permit essentially
identical methods to be applied in many different situations. This is why shape
dynamics is a universal framework. 

\subsection{Newtonian dynamics in shape space}

We now identify the role that absolute space and time play in
Newtonian dynamics by \emph{projection to shape space}.
We have
removed structure from the $q$'s in $\textrm {Q}^N$, reducing them to points
$s\in\textrm Q_{ss}^N$. This is projection of $q$'s. We can also project
complete Newtonian histories $q(t)$. To include time at the start, we adjoin to
$\textrm Q^N$ the space $\textrm T$ of absolute times $t$, obtaining the space
$\textrm Q^N\textrm T$. Newtonian histories are then (monotonically rising)
continuous curves in $\textrm Q^N\textrm T$. However, clocks are parts of the
universe; there is no external clock available to provide the
reading for the $\rm T$ axis. All the objective information is carried by the
successive configurations of the universe. We must therefore remove the $\rm T$
axis and, in the first projection, label the points representing the
configurations in $Q^N$ by an arbitrary increasing parameter $\lambda$ and then
make the further projection to the shape space  $\textrm
Q_{ss}^N$. The history becomes $s(\lambda)$ (Fig.~2).  A history is the next
most fundamental concept in shape dynamics. 
There is no `moving now' in this concept. History is not a
spot moving along $s(\lambda)$, lighting up `nows' as
it goes. It is the curve;
$\lambda$ merely labels its points. Newtonian dynamics being time-reversal
invariant, there is no past-to-future direction on curves in $\textrm Q_{ss}^N$.

\begin{figure}
 \begin{center}
 \includegraphics[width=11cm]{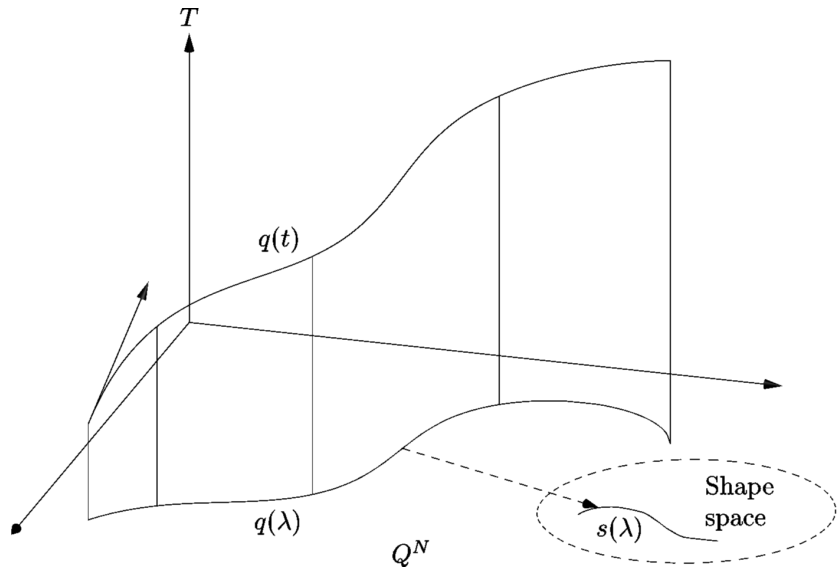}

\caption{In Newtonian dynamics, the history of a system
is a monotonically rising curve $q(t)$ in $Q^NT$ or a curve $q(\lambda)$ in
$Q^N$ labelled by a monotonic $\lambda$.
The objective observable history is the projected curve $s(\lambda)$ in shape
space $\textrm Q_{ss}^N$.}

 \end{center}

\end{figure}

Given a history of shapes $s$, we can define a shape \emph{velocity}.
Suppose first that in fact by some means we can define
a distinguished parameter $p$, or
\emph{independent variable}, along a suitably continuous curve in $\textrm
Q_{ss}^N$.
Then at each point along the curve we have a shape $s$ and its
(multi-component) velocity $\rm ds/\rm dp$. This is a \emph{tangent vector} to
the curve. If we have no $p$ but only an arbitrary
$\lambda$, we can still define shape velocities $\rm ds/\rm d\lambda = s'$,
but all we really have is the direction $d$ (in $\textrm Q_{ss}^N$)
in which $s$ is changing. The difference between tangent vectors and directions
associated with curves in shape space will be important later.

We can now identify the mismatch that, when
eliminated, leads to the shape-dynamic ideal. To this
end, we recall Laplacian determinism in Newtonian
dynamics: given $q$ and $\dot q$ at some instant, the
evolution of the system is uniquely determined (the particle masses and the
force law assumed known). The question is this: given
the corresponding shape projections $s$ and $d$, is
the evolution \emph{in shape space} $\textrm Q_{ss}^N$ uniquely determined? The
answer is no for a purely geometrical reason. The fact is that certain initial
velocities which are objectively significant in Newtonian
dynamics can be generated by purely group actions.
To be precise, different Newtonian velocities can be generated from
identical data in $\textrm Q_{ss}^N$. This is illustrated for the 3-body problem
(in two dimensions) in Fig.~3.

\begin{figure}
 \begin{center}
 \includegraphics[width=11cm]{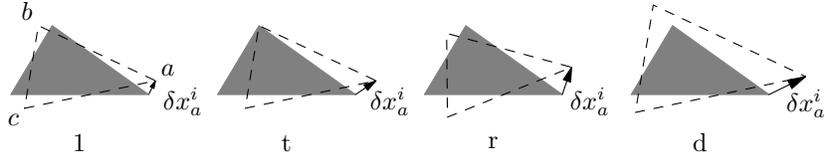}

\caption{The two triangles of slightly different shapes formed by three
particles $a, b, c$ define a point $s$ and direction $d$ \emph{uniquely} in
shape
space, but changes to an original placing (1) in Euclidean space of the dashed
triangle relative to the grey one generated by translations (t), rotations
(r), and dilatations (d) give rise to different Newtonian
initial velocities $\rm dx_a^i/\rm dt$.}

 \end{center}

\end{figure}

I will not go into the details of the proof (see
\cite{Saari2005}), but in a Newtonian $N$-body system the velocities at any
given instant can be
uniquely decomposed into parts due to an intrinsic change of shape and three
further parts due to the three different group actions -- translations,
rotations, and dilatations -- applied as in Fig.~3.
These actions are obviously `invisible' in the shape-space
$s$ and $d$, which define only the shape
and the way it is changing.

By Galilean relativity, translations of the system have no effect in
$\textrm Q_{ss}^N$. We can
ignore them but not rotations and dilatations. Four
dimensionless
dynamically effective quantities are associated with them. First, two angles
determine the direction in space of a rotation axis.
Second, from the kinetic energies
associated with rotation, $T_r$, dilatation $T_d$, and change of shape, $T_s$,
we can form two dimensionless ratios, which it is natural to
take to be $T_r/T_s$ and $T_d/T_s$ (since change of shape, represented by
$T_s$, is our `gold
standard'). Thus, the kinematic action of the Lie groups generates
\emph{four} parameters that affect the histories in shape space without
changing the initial $s$ and $d$. This is already so for pure inertial motion.
If forces are present, there is a fifth parameter, the ratio $T/V$ of the
system's kinetic energy $T$ to its potential energy $V$, that is dynamically
significant but is also invisible in the $s$ and $d$ in shape space.

We now see that although Newtonian dynamics seems wonderfully rational and
transparent when expressed in an inertial frame of reference, it does not
possess perfect
Laplacian determinism in
shape space. This failure appears especially odd if
$N$ is large. Choose some coordinates $s_i, i=1,2,...,3N-7,$ in shape
space and take one of them,
call it $\tau$, as a surrogate for Newton's $t$. If
only shapes had dynamical
effect, then by analogy with inertial-frame
Newtonian dynamics, the initial values of $\tau,
s_i, \textrm ds_i/\textrm d\tau, i=1..., 3N-8$ would fix the evolution. They do
not. 

Five more
data are needed and must be taken from
among the second derivatives $ \textrm d^2s_i/\textrm d\tau^2$.
Moreover, no matter how large $N$, say a million as in a globular
cluster, we always
need just five.\footnote{If $N=3~ \textrm{or}~ 4$, there are
insufficient $ \textrm d^2s_i/\textrm d\tau^2$'s and we need higher
derivatives too.} They make no sense from the shape-dynamic perspective.
Poincar\'e,
writing as a philosopher deeply committed to relationalism, found the need for
them repugnant \cite{Poincar'e1902, Poincar'e1905}. But, in
the face
of the manifest presence of angular momentum in the solar system, he resigned
himself to the fact that there is more to dynamics than, literally, meets the
eye.\footnote{Poincar\'e's penetrating analysis, on which this subsection is
based, only takes into account the role of angular momentum in the `failure' of
Newtonian dynamics when expressed in relational quantities. Despite its
precision and clarity, it has been almost totally ignored in the discussion of
the absolute vs relative debate in dynamics.} In fact, the extra
$ \textrm d^2s_i/\textrm d\tau^2$'s \emph{are} explained by
Newton's assumption of an all-controlling but
invisible frame for dynamics. They are the evidence, and the sole evidence at
that, for absolute space.

For some reason, Poincar\'e did not consider Mach's suggestion
\cite{Mach1883, Mach1960} that the
universe in its totality might somehow determine the structure
of the dynamics observed locally. Indeed, the universe exhibits evidence for
angular momentum in innumerable localized systems but none overall. This
suggests that, regarded as a
closed dynamical system, it has no angular momentum and meets the
\emph{Poincar\'e principle}: either a point and direction (strong form) or a
point and a tangent vector (weak form) in the universe's
shape space determine its evolution. The stronger form of the principle will
hold if the universe satisfies a \emph{geodesic principle} in $Q_{ss}^N$, since
a point and a direction are the initial conditions for a geodesic. The need for
the two options, either of which may serve as the definition of Mach's principle
\cite{Barbour2010}, will be clarified in the next section.

To summarize: on the basis of Poincar\'e's analysis and intuition, we would like
the universe to satisfy Laplacian determinism in its shape space and not merely
in a special frame of reference that employs kinematic structure not present in
shape space. 

\section{The Universal Structure of Shape Dynamics}

\subsection{The elimination of time \label{jacobi}}

In standard dynamical theory, the time $t$ is an independent
variable supplied by an external clock. But any clock is a
mechanical system. If we wish to treat the universe as a single
system, the issue of what clock, if any, to use becomes critical. In fact, it
is not necessary to use any clock.

This can be demonstrated already in
$Q^N$. We simply proceed without a clock. Histories of the system are
then just curves in $Q^N$, and we seek a law that determines 
them. An obvious possibility is to
define a metric on $Q^N$ and require histories to
be geodesics with respect to it. 

A metric is readily found because the Euclidean geometry of space that defines
$Q^N$ in the first place also defines a natural metric on $Q^N$:
\begin{equation}
 \textrm ds_{kin}
=\sqrt{\sum_a{m_a\over 2}\rm d\textbf{x}_a\cdot\rm d\textbf{x}_a}.\label{metric}
\end{equation}
This is called the \emph{kinetic metric} \cite{Lanczos1949}; division of
$\rm d\textbf{x}_a$
by an external $\rm dt$ transforms the radicand into the Newtonian kinetic
energy. We may call (\ref{metric}) a
\emph{supermetric}. We shall see how it enables us
to exploit structure defined at the level of $\rm Q^N$ at the
shape-space level. 

We can generate further such supermetrics from (\ref{metric}) by multiplying its
radicand
by a function on $Q^N$, for example
$\sum_{a<b}m_am_b/r_{ab}$. We obtain a whole family of geodesic principles
defined by the variational requirement
\begin{equation}
\delta I=0,~I=2\int\textrm
d\lambda\sqrt{(E-V(q))T_{kin}},~T_{kin}:={1\over 2}\sum_am_a{\textrm
d\textbf{x}_a\over\textrm d\lambda}\cdot{\textrm
d\textbf{x}_a\over\textrm d\lambda},\label{metric1}
\end{equation}
where $\lambda$ is a curve parameter, the $2$ is for convenience
and, since a constant is a function on $Q^N$, the constant $E$ reflects its
possible presence.

The Euler--Lagrange equations that follow from (\ref{metric1}) are
\begin{equation}
 {\textrm d\over\textrm d\lambda}\left(\sqrt{{E-V}\over
T_{kin}}\,m_a{\textrm d\textbf{x}\over\textrm
d\lambda}\right)=-\sqrt{T_{kin}\over{E-V}}{\partial
V\over\partial\textbf{x}_a}.\label{EL}
\end{equation}
This equation simplifies if we choose the
freely specifiable $\lambda$ such that
\begin{equation}
 E-V=T_{kin}.\label{newt}
\end{equation}
If we denote this $\lambda$ by $t$, then (\ref{EL}) becomes Newton's second law
and (\ref{newt}) becomes the energy theorem. However, in our initially timeless
context it becomes the definition of an emergent time, or better
\emph{duration}, created by a geodesic principle. In fact, the entire objective
content of Newtonian dynamics for a closed system is recovered. It is
illuminating to give the explicit expression for the increment of this emergent
duration:
\begin{equation}
 \delta
t=\sqrt{\sum_am_a\delta\textbf{x}_a\cdot\delta\textbf{x}_a\over{2(E-V)}}.
\label{dist}
\end{equation}
This is the first example of the holism of relational dynamics: the time that 
we take to flow locally everywhere is a distillation of all the changes
everywhere in the universe. Since everything in the universe interacts with
everything else, every difference must be taken into account to obtain the
\emph{exact} measure of time. The universe is its own clock.

The definition of duration through (\ref{dist}) is unique (up to origin and
unit) if clocks are to have any utility. Since we
use them to keep appointments, they are useless unless they
march in step. This leads unambiguously to (\ref{dist}) as the only sensible
definition. For suppose an island universe contains within it
subsystems that are isolated in the Newtonian sense. We want to
use the motions within each to generate a time signal. The resulting signals
must all march in step with each other. Now this will happen if, for each
system, the signal is generated using (\ref{dist}). The reason is
important. Suppose we used only the numerators in
(\ref{dist}) to measure time; then subsystems without interactions would
generate time signals
that march in step, but with
interactions one system may be sinking
into its potential well as another is rising out of its. Then the `time'
generated by the former will pass faster than the latter's. However the
denominators in (\ref{dist}) correct this automatically since
$E-V$ increases or decreases with $T$. Time must be
measured by some motion, but for generic systems only the time label that
ensures conservation of the energy can meet the marching-in-step
criterion. Duration is defined as uniquely as
entropy is through the logarithm of probability.

In textbooks, (\ref{metric1}) is derived as  Jacobi's principle
\cite{Lanczos1949} and
used to determine the dynamical orbit of systems in $Q^N$ (as, for example, a
planet's orbit, which is not to be confused with a group orbit). The speed
in
orbit is then determined from (\ref{newt}) regarded as the energy theorem. The
derivation above provides the deeper interpretation of (\ref{newt}) in a closed
system. It is the definition of time. Note that time is
eliminated from the initial kinematics by \emph{a square root} in the
Lagrangian. This
pattern will be repeated
in more refined relational settings below, in which we can address the question
of what potentials $V$ are allowed in relational dynamics.

A final comment. Time has always appeared elusive. It is represented in
dynamics as the real line $R^1$. Instants are mere points on the
line, each identical to the other. This violates the principle that things can
be distinguished only by differences. There must be variety. In
relational dynamics $R^1$ is redundant and there are only
configurations, but they double as instants of time. The need for
variety is met.

\subsection{Best matching}

The next step is to determine curves in shape space
$Q_{ss}^N$ that satisfy the strong or weak form of the Poincar\'e principle.
As already noted, the strong form, with which we begin, will be satisfied by
geodesics with respect to a metric defined on $Q_{ss}^N$. For this, given two
nearly identical shapes,
$s_1,s_2,$ i.e., neighbouring points in $Q_{ss}^N$, we need to define a
`distance' between them based on their difference and nothing else. Once
again we use the Euclidean geometry that underlies both $Q^N$ and $Q_{ss}^N$.

Shape $s_1$ in $\textrm Q_{ss}^N$ has infinitely many representations in
$Q^{N}$: all of
 the points on its group orbit in $Q^{N}$. Pick one with coordinates 
$\textbf{x}_a^1$. Pick a nearby point on the orbit of $s_2$ with coordinates
$\textbf{x}_a^2$. In Newtonian dynamics, the coordinate differences 
$\rm d\textbf{x}_a=\textbf{x}_a^2-\textbf{x}_a^1$ are physical displacements,
but in shape dynamics they mix physical difference of shape with
spurious difference due to the arbitrary positioning of $s_1$ and $s_2$ on their
orbits. To obtain a measure of the shape difference, hold $s_1$ fixed in
$Q^{N}$ and move $s_2$ around in its orbit,\footnote{Recall that a group orbit
is 
generically a multi-dimensional space.} for the moment using only Euclidean
translations and rotations. This changes
$\rm d\textbf{x}_a=\textbf{x}_a^2-\textbf{x}_a^1$ and simultaneously
\be
\rm ds_{trial}:=\sqrt{(E-V)\sum_a
{m_a\over 2}\rm d\textbf{x}_a\cdot\rm d\textbf{x}_a}.\label{trial}
\ee

Since (\ref{trial}) is positive definite and defines a nonsingular metric on
$Q^{N}$, it will be possible to move shape $s_2$ into the unique position in its
orbit at which (\ref{trial}) is minimized  (for
given position of $s_1$). This
unique position can be characterized in two equivalent ways: 1) Shape $s_2$ has
been moved to the position in which it most closely
`covers' $s_1$, i.e., the two shapes, which are incongruent, have been brought
as close as possible to congruence, as measured by (\ref{trial}). This is the
\emph{best-matched} position. 2) The $3N$-dimensional vector joining $s_1$ and
$s_2$ in their orbits in $Q^{N}$ is \emph{orthogonal} to the orbits. This is
true in the first place for the kinetic metric, for which
$E-V=1$, but also for all choices of $E-V$. For each, the best-matched
position is the same but there is
a different best-matched `distance' between $s_1$ and $s_2$:
\be
\rm ds_{bm}:=\textrm{min~of~}\sqrt{(E-V)\sum_a
{m_a\over 2}\rm d\textbf{x}_a\cdot\rm d\textbf{x}_a}~\textrm{~between~orbits}
.\label {bm}
\ee

Because orthogonality of two vectors can only be established if all components
of both vectors are known, best matching introduces a further degree of holism
into relational physics. The two ways of conceptualizing best
matching are shown in Fig.~4 for the 3-body problem in two dimensions.

\begin{figure}
 \begin{center}
 \includegraphics[width=11cm]{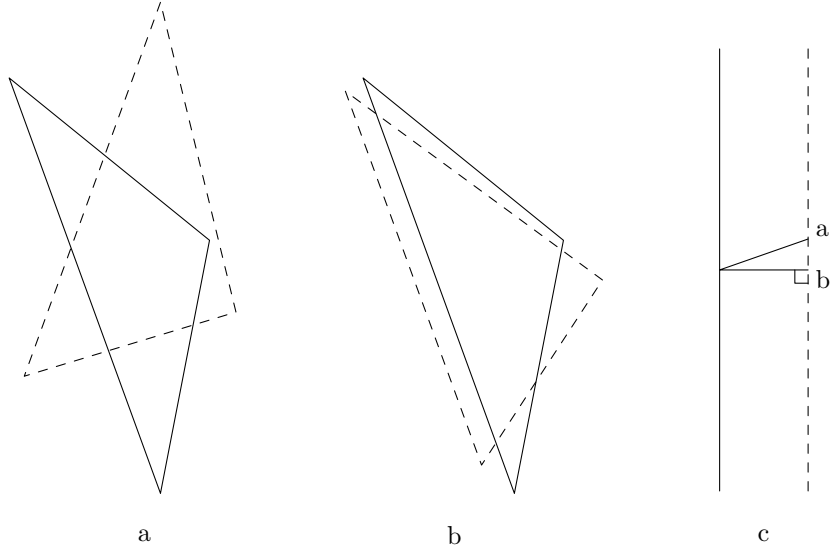}

\caption{$a$) An arbitrary placing of the dashed triangle
relative to the undashed triangle; b) the best-matched placing reached
by translational and rotational minimization of (\ref{trial}); $c$) the two
positions of the triangle configurations on their group orbits in $Q^N$. The
connecting `strut' is orthogonal with respect to the supermetric on $Q^N$ in the
best-matched position. Best matching brings the centres of mass to coincidence
and reduces the net rotation to zero.}

 \end{center}

\end{figure}

It is important that the orthogonal
separation (\ref{bm}) is the same at all
points within the orbits of either $s_1$ or $s_2$. This is because the metric
(\ref{trial})  on $Q^{N}$ is \emph{equivariant}: if the same group
transformations
are applied to the
configurations in $Q^{N}$ that represent $s_1$ and $s_2$, the value of
(\ref{trial}) is
unchanged. In differential-geometric
terms, equivariance is present because the translation and rotation
group orbits are Killing vectors of the kinetic metric in $Q^{N}$. The
equivariance
property only holds if $E-V$ satisfies definite conditions, which I have tacitly
assumed so far but shall spell out soon. 

In fact, it is already lost if we
attempt to include dilational best matching with respect to the kinetic metric
in order to determine a `distance' between shapes rather than only relative
configurations as hitherto. For suppose we represent two shapes by
configurations of given sizes in $Q^{N}$ and find their best-matched separation
$\textrm d_{bm}$ using Euclidean translations and rotations. We obtain some
value for \bm. If we now change the scale of one of the shapes, 
 \bm\, must change because the kinetic metric
has dimensions $m^{1/2}l$ and scales too. To correct for this in a natural way,
we can divide the kinetic metric by the square root of $I_{cms}$, the
centre-of-mass
moment of inertia (\ref{cms}), and then best match to get the inter-shape
distance
\be
\rm ds_{sbm}:=\textrm{min~of~}\sqrt{I_{cms}^{-1}\sum_a
m_a\rm d\textbf{x}_a\cdot\rm d\textbf{x}_a}\textrm{~between~orbits}
.\label {sbm}
\ee
As it must be, $\rm ds_{sbm}$ is dimensionless and defines a metric on
$\textrm Q_{ss}^N$.  It is precisely such a metric that we need in order to
implement Poincar\'e's principle.

Terminologically, it will be convenient to call directions in $Q^{N}$ that lie
entirely in group orbits \emph{vertical} and the best-matched orthogonal
directions \emph{horizontal}. Readers familiar with fibre bundles
will recognize
this terminology. A paper presenting best-matching theory \emph{ab
initio} in terms of fibre bundles is in preparation. 

\subsection{The best-matched action principle}

We can now implement the strong Poincar\'e principle. We calculate
in $Q^{N}$, but the reality unfolds in $\textrm Q_{ss}^N$. The task is this:
given two shapes $s_a$
and $s_b$ in $\textrm Q_{ss}^N$, find the geodesic that joins them. The distance
along the
trial curves between $s_a$
and $s_b$ is to be calculated using the best-matched metric
(\ref{sbm}) found in $Q^{N}$ and then `projected' down to $\textrm Q_{ss}^N$.
The
projected metric is unique because the best-matching metric in $Q^{N}$ is
equivariant.

The action principle in $Q^{N}$ has the form
\be
\delta I_{bm}=0,~I_{bm}=2\int\rm d\lambda\sqrt{WT_{bm}},~T_{bm}={1\over
2}\sum_a{\rm
d\bf{x}_a^{bm}\over\rm d\lambda}\cdot{\rm
d\bf{x}_a^{bm}\over\rm d\lambda},\label{bma}
\ee
where $\rm d\bf{x}_a^{bm}/\rm d\lambda$ is the limit of
$\delta\bf{x}_a^{bm}/\delta\lambda$ as $\rm d\lambda\rightarrow 0$, and the
potential-type term $W$ must be such that equivariance holds.
In writing the action in this way, I have taken a short cut. Expressed properly
\cite{Barbour1982},
$I_{bm}$ contains the generators of the various group transformations,
and the variation with respect to them leads to the best-matched velocities 
$\rm d\bf{x}_a^{bm}/\rm d\lambda$. It
is assumed in (\ref{bma}) that this variation has already been done. 

The action (\ref{bma}) is interpreted as follows. One first fixes a
trial curve in $\textrm Q_{ss}^N$ between $s_a$
and $s_b$ and represents it by a trial
curve in $Q^{N}$ through the orbits of the shapes in the
$\textrm Q_{ss}^N$ trial curve. The $Q^{N}$ trial curve must never `run
vertically'. It may run
orthogonally to the orbits,
and this is just what we want. For if it does, the $\delta\bf{x}_a$ that
connect the orbits are best
matched. It is these $\delta\bf{x}_a^{bm}$, dependent only on the
shape differences, that are to determine the action. 

To make the trial curve in
$Q^{N}$ orthogonal, we divide it into infinitesimal
segments between adjacent orbits $1,2,3,...,m$
(orbits $1$ and $m$ are $s_a$ and $s_b$, respectively). We hold the
initial point of segment 1--2 fixed and move the other end
into the horizontal best-matched position on orbit 2. We then move the
original 2--3 segment into the horizontal with its
end 2 coincident with the end of the adjusted 1--2 segment. We do this all the
way to the
$s_b$ orbit. Making the segment lengths tend to zero, we obtain a
smooth horizontal curve. Because the $Q^{N}$ metric is equivariant, this curve
is
not unique -- its initial point can be moved `vertically' to any other point
on the initial orbit; all the other points on the curve are then moved
vertically by the same amount. If $M$ is the dimension of the best-matching
group ($M=7$ for the similarity group), we obtain an
$M$-parameter family of horizontal best-matched curves that all yield a
common unique value for the action along the trial curve in $\textrm Q_{ss}^N$.
This is illustrated in Fig.~5.

\begin{figure}
 \begin{center}
 \includegraphics[width=11cm]{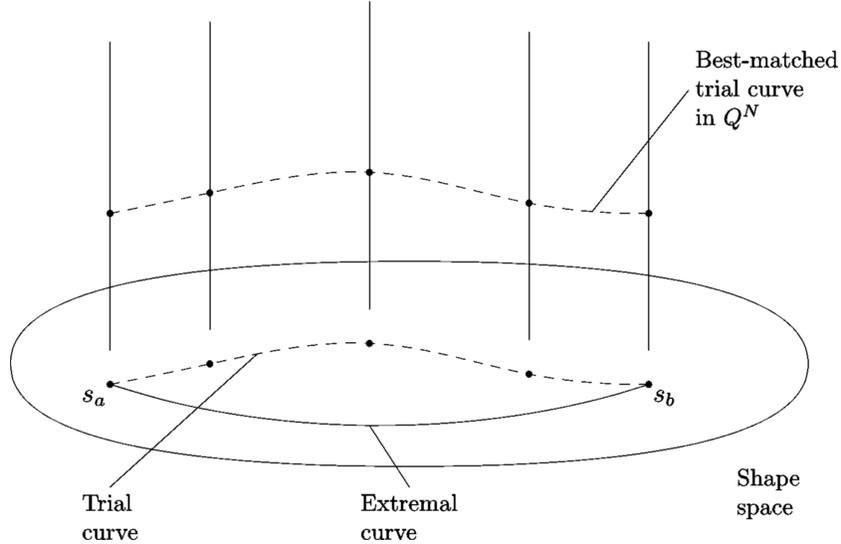}

\caption{The action associated with each trial curve between shapes $s_a$
and $s_b$ in shape space is calculated by finding a best-matched curve in $Q^N$
that runs through the group orbits `above' the trial curve.
The best-matched curve is determined uniquely apart from `vertical lifting' by
the same amount in each orbit, which does not change the best-matched action
$I_{bm}$. The trial curve in shape space for which $I_{bm}$ is extremal is the
desired curve in shape space.}

 \end{center}

\end{figure}

In this way we obtain the action for all trial curves in shape
space between  $s_a$ and $s_b$. The best-matching construction
ensures that the action depends only on the shapes that are explored by the
trial curves and nothing else. It remains to find which
trial curve yields the shortest distance between $s_a$ and $s_b$. This requires
us to vary the trial curve in $\textrm Q_{ss}^N$, which of course changes the
associated trial
curves in $Q^{N}$, which, when best matched, give different values for the
best-matched action. When we find the (in general unique) curve for which the
shape-space action is stationary, we have found the solution that satisfies the
strong Poincar\'e principle. Theories satisfying only the weak principle arise
when the equivariance condition imposed on $W$ in (\ref{bma}) is somewhat
relaxed, as we shall now see.

\subsection{Best-matching constraints and consistency}

To obtain a definite representation in the above picture of best matching, we
must refer the initial shape $s_1$ to a
particular Cartesian coordinate system with a definite choice of scale.
This `places' shape $s_1$ at some position on its group orbit in $Q^N$.
If  we now place the next, nearly
identical shape $s_2$ on its orbit close to the position chosen for $s_1$
on its orbit but not in the best-matched position, we
obtain certain coordinate differences
$\delta\textbf{x}_a=\textbf{x}_a^2-\textbf{x}_a^1$. Dividing these by a nominal
$\delta t$, we obtain velocities from which, in Newtonian
terms, we can calculate a total momentum $\bf P=\sum_am_a\dot\bf{x}_a$, angular
momentum $\bf L=\sum_am_a\bf{x}_a\times\dot\bf{x}_a$ and rate of change of 
$I$: $\dot I
= D= 2\sum_am_a\bf{x}_a\cdot\dot\bf{x}_a$. We can change their values by acting
on $s_2$ with
translations, rotations, and dilatations respectively. Indeed, it is intuitively
obvious
that by choosing these group transformations appropriately we can ensure that
\be
\bf P=0,\label{P}
\ee
\be
\bf L=0,\label{L}
\ee
\be
D=0.\label{D}
\ee
It is also intuitively obvious that the fulfilment of these conditions is
precisely the indication that the best-matching position has been reached.

Let us now stand back and take an overall view. \emph{The} reality in shape
dynamics is simply a curve in $\textrm Q_{ss}^N$, which we can imagine traversed
in either
direction. There is no rate of change of shapes,
just their succession. The only convenient way to
represent this succession is in $Q^{N}$. However, any one curve in $\textrm
Q_{ss}^N$,
denote it $C_{ss}$, is represented by an infinite set $\{C_{ss}^{Q^N}\}$ of
curves in $Q^{N}$. They all pass through the orbits of the shapes in
$C_{ss}$, within which the
$\{C_{ss}^{Q^N}\}$ curves can run anywhere. Prior to the introduction of the
best-matching dynamics, all the curves $\{C_{ss}^{Q^N}\}$ are
equivalent representations of $C_{ss}$ and no curve parametrization is
privileged.

Best matching changes this by singling out curves in the set
$\{C_{ss}^{Q^N}\}$ that `run horizontally'. They are \emph{distinguished
representations}, uniquely determined by the best matching
up to a seven-parameter freedom of position in one nominally chosen initial
shape in its orbit. There is also a distinguished curve parametrization 
(Sec.~\ref{jacobi}), uniquely fixed up to its origin and unit. When
speaking of the distinguished representation, I shall henceforth mean that
the curves in $Q^{N}$ and their parametrization have both been chosen in the
distinguished form (modulo the residual freedoms). 

Let us now consider how the dynamics that actually unfolds in $\textrm Q_{ss}^N$
is seen to
unfold in the distinguished representation. From the form of the action
(\ref{bma}), knowing that Newton's second law can be recovered from Jacobi's
principle by choosing the distinguished curve parameter using (\ref{dist}), we
see
that we shall recover Newton's second law exactly. We derive not only Newton's
dynamics but also the frame and time in which it holds (Fig.~6). There is a
further  
bonus, for the best-matching dynamics is more predictive: the conditions
(\ref{P}), (\ref{L}) and (\ref{D}) must hold at any initial point that we choose
and be maintained subsequently. Such conditions
that depend only on the initial data (but not accelerations) and must be
maintained (propagated) are called \emph{constraints}. This is
the important topic treated by Dirac \cite{Dirac1964}.

\begin{figure}
 \begin{center}
 \includegraphics[width=11cm]{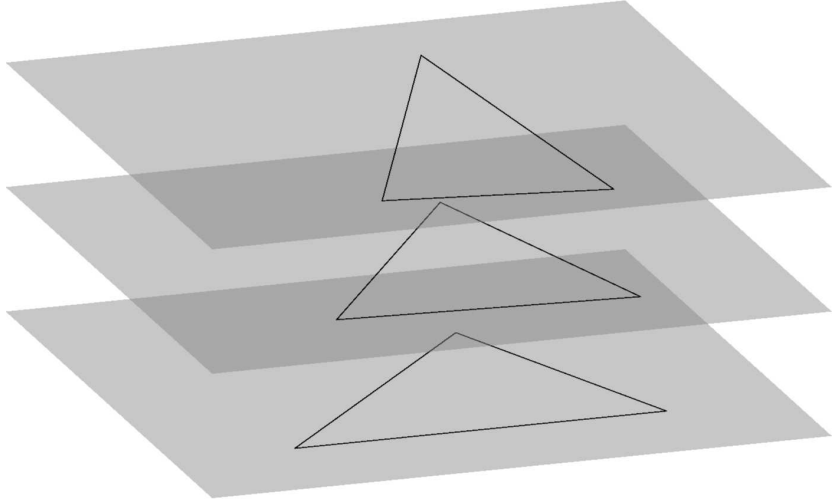}

\caption{The distinguished representation of best-matched shape dynamics for
the 3-body problem. For the initial shape, one chooses an arbitrary position in
Euclidean space. Each successive shape is placed on its predecessor in the
best-matched position (`horizontal stacking'). The `vertical' separation is
chosen in accordance with distinguished curve parameter $t$ determined by the
condition (\ref{dist}). In the framework thus created, the particles behave
exactly as Newtonian particles in an inertial frame of reference with total
momentum and angular momentum zero.}

 \end{center}

\end{figure}

Since the dynamics in the distinguished representation is governed by Newton's
second law, we need to establish the conditions under which it will propagate
the constraints (\ref{P}), (\ref{L}) and (\ref{D}). In fact, we have to impose
conditions on the potential term $W$ in (\ref{bma}). If
(\ref{P}) is to propagate, $W$ must be a function of the
coordinate differences $\bf x_a -\bf x_b$; if (\ref{L}) is to propagate, $W$ can
depend on only the inter-particle separations $r_{ab}$. These are
both standard conditions in Newtonian dynamics, in which they are usually
attributed to the homogeneity and isotropy of space. Here they ensure
consistency of best matching wrt the Euclidean group. Propagation of (\ref{D})
introduces a novel element. It requires $W$ to be homogeneous with length
dimension $l^{-2}$. This
requirement is immediately obvious in (\ref{bma}) from the length dimension
$l^{2}$ of the kinetic term, which the potential must balance out. Note that in
this case a constant $E$, corresponding to a nonzero energy of the system,
cannot appear in $W$. The system must, in Newtonian terms, have total
energy zero. However, potentials with
dimension $l^{-2}$ are virtually never considered in Newtonian dynamics because
they do not appear to be realized in nature.\footnote{It is in fact possible to
recover Newtonian gravitational and electrostatic forces exactly from $l^{-2}$
potentials by dividing the $l^{-1}$ Newtonian potentials by the square root of
the moment
of inertia $I_{cms}$. This is because $I_{cms}$ is dynamically conserved and is
effectively
absorbed into the gravitational constant G and charge values. However, the
presence of $I_{cms}$ in the action leads to an additional force that has the
form of a
time-dependent `cosmological constant' and ensures that $I_{cms}$ remains
constant.
See \cite{Barbour2003} for details.} I shall discuss this issue in the
next subsection after some general remarks.

Best matching is a process that determines a metric on
$\textrm Q_{ss}^N$. For this, three things are
needed: a supermetric on $Q^{N}$, best matching to find the
orthogonal inter-orbit separations determined by it, and the
equivariance property that ensures identity of them at all positions on the
orbits. Nature gives us the metric of Euclidean
space, and hence the supermetric on $Q^N$;
the second and third requirements
arise from the desire to implement Poincar\'e type dynamics in $\textrm
Q_{ss}^N$. The orbit
orthogonality, leading to the constraints (\ref{P}), (\ref{L}), and (\ref{D}),
distinguishes best-matched dynamics from Newtonian theory, which
imposes no such requirements. Moreover, the constraint propagation needed for
consistency of best matching enforces symmetries of the potential that in
Newtonian theory have to be taken as facts additional to the basic structure of
the theory.

It is important that the constraints  (\ref{P}), (\ref{L}), and (\ref{D}) apply
only to the `island
universe' of the complete $N$-body system.
Subsystems within it that are isolated from each other,
i.e., exert negligible forces on each other, can perfectly well have
nonvanishing values of $\bf P, \bf L, D$. It is merely necessary that their
values for all of the subsystems add up to zero. However, the consistency
conditions imposed on the form of the potential must be maintained at the level
of the subsystems.

We see that shape dynamics has several advantages over
Newtonian dynamics. The two forms of dynamics have Euclidean
space in common, but
shape dynamics derives all of Newton's additional kinematic structure: 
absolute space (inertial frame of reference), the metric of time (duration),
and the symmetries of the potential. Besides these
qualitative advantages, shape dynamics is more powerful: fewer initial data 
predict the evolution. 

This subsection has primarily been concerned with the problem of defining
change of position. Newton clearly understood
that this requires
one to know when one can say a body is \emph{at the same place} at different
instants of time. Formally at least he solved this problem by the notion of
absolute space. Best matching is \emph{the relational alternative to absolute
space}. For when one configuration has been placed relative to another in the
best-matched position, every position in one configuration is uniquely paired
with a position in the other. If a body is at these paired positions at the two
instants, one can say it is at the same place. The two positions are
\emph{equilocal}. The image of `placing' one configuration on another in the
best-matched position is clearly more intuitive than the notion of inter-orbit
orthogonality. It makes the achievement of relational equilocality manifest. It
is also worth noting that the very thing that creates the problem of defining
change of position -- the action of the similarity group -- is used to resolve
it in best matching.

\subsection{Two forms of scale invariance}

We now return to the reasons for the failure of Laplacian
determinism of Newtonian dynamics when considered in
shape space. This will explain why it is desirable to keep open the option of
 the weaker form of the Poincar\'e principle. It will be helpful to consider
Newtionian dynamics once more in the form of Jacobi's principle:
\be
\delta I=0,~I=2\int\textrm
d\lambda\sqrt{(E-V(q))T_{kin}},~T_{kin}:={1\over 2}\sum_am_a{\textrm
d\textbf{x}_a\over\textrm d\lambda}\cdot{\textrm
d\textbf{x}_a\over\textrm d\lambda}.\label{metric2}
\ee
Typically $V(q)$ is a sum of terms with different, usually integer
homogeneity degrees: gravitational and electrostatic
potentials have $l^{-1}$, harmonic-oscillator potentials are $l^2$.
Moreover, since (\ref{metric2}) is timeless and
only the dimensionless mass ratios have objective meaning, length is the sole
significant dimension. Because all terms in $V$ must have
the same dimension, dimensionful
coupling constants must appear. One can be set to unity
because an overall factor multiplying the action has no effect on its
extremals. If we take G=1, a fairly general action will have 
\be
W=E-V=E+\sum_{i<j}{m_im_j\over r_{ij}}-g_iV_i,\label{pot}
\ee
where the $V_i$'s have different homogeneity degrees, some perhaps the
same (as for gravity and electrostatics). Now the crux: different
$E$ and $g_i$ values lead to different curves in shape space, but not to any
differences that can be expressed through an initial point and direction in
$\textrm Q_{ss}^N$, which cannot encode dimensionful information.
Thus, each
such $E$ and $g_i$ present in (\ref{pot}) adds a one-parameter degree
of uncertainty into the evolution from an initial point and direction in
$\textrm Q_{ss}^N$. If
the strong
Poincar\'e principle holds, this unpredictability is
eliminated. There may still be several different terms in $V$ but they must all
have the same homogeneity degree $-2$ and dimensionless coupling constants; in
particular, the constant $E$ cannot
be present. Note also that \emph{any} best matching enhances predictability
and eliminates potentially redundant structure. But other factors may count.
Nature may have reasons not to best match with respect to all conceivable
symmetries.

Indeed, the foundation of particle shape dynamics on the
similarity group precluded consideration of the
larger general linear group. I suspect that this group would leave too little
structure to construct dynamics at all easily and that angles are the
irreducible minimum needed. Another factor, possibly more relevant,
is the difference between velocities (and
momenta), which are vectors, and directions, which are not (since
multiplication of them by a number is meaningless). Vectors and vector
spaces have
mathematically desirable properties. In quantum mechanics, the vector nature of
momenta ensures that the momentum and configuration spaces have the same number
of dimensions, which is important for the equivalence of the position and
momentum representations (transformation theory). If we
insist on the strong Poincar\'e principle, the equivalence will be lost for a
closed
system regarded as an island universe. There are then two possibilities: either
equivalence is lost, and transformation theory only arises for subsystems (just
as inertial
frames of reference arise from shape dynamics), or the strong Poincar\'e
principle is relaxed just enough to maintain equivalence.

There is an interesting way to do this. In the generic
$N$-body problem the energy $E$ and angular
momentum $L$, as dimensionful quantities, are not scale invariant. But they
are if $E=L=0$. Then the behaviour is scale invariant. Further
\cite{Barbour2003}, there is
a famous qualitative result in the $N$-body problem, first proved by Lagrange,
which is that  $\ddot I>0$ if $E\ge 0$. Then the curve of $I$
as a function of time is concave upwards and its time derivative, which is
$2D$ (defined just before (\ref{D})), is strictly monotonic, increasing from
$-\infty$ to $\infty$ (if the
evolution is taken nominally to begin at
$D=-\infty$). 

Now suppose that, as I conjecture, in its classical limit
the quantum mechanics of the universe does require there to be velocities (and
with them momenta) in shape space and its geometrodynamical generalization, to
which
we come soon. Then there must at the least be a one-parameter family of
solutions that emanate from a point and a direction in $\textrm Q_{ss}^N$. We
will certainly want rotational best
matching to enforce $L=0$. We will then have to relax dilatational best
matching in such a way that a one-parameter freedom is introduced. In the
$N$-body problem we can do this, without having a best-matching symmetry
argument that enforces it, by requiring $E=0$. The
corresponding one-parameter freedom in effect
converts a direction in $\textrm Q_{ss}^N$ into a vector. The interesting thing
is now that, by Lagrange's result, $D$ is monotonic
when $E=0$. This means that the shape-space dynamics can be monotonically
parametrized by the dimensionless ratio $D_c/D_0$, where $D_0$ is an initial
value of $D$ and $D_c$ is the current value. Thus, $D_c/D_0$ provides
an objective
`time' difference between shapes $s_1$ and $s_2$. The scare quotes are used
because it does not march in step
with the time defined by (\ref{newt}). 

An alternative dimensionless parametrization of the shape-space curves in this
case is by means of the (not necessarily monotonic) ratio $I_c/I_0$. Because the
moment of inertia measures the `size' of the universe, this ratio measures `the
expansion of the universe' from an initial size to its current size. One might
question whether in this case one should say that the dynamics unfolds on shape
space. Size still has some meaning, though not at any one instant but only as a
ratio at two instants. Moreover, on shape space this ratio plays the
role of `time' or `independent variable'. It does not appear as a dependent
dynamical variable. This is related to the cosmological puzzle that I
highlighted at
the end of the introduction: from the shape-dynamic perspective, the
expansion of the universe seems to be made possible by a last vestige of
Newton's absolute space. I shall return to this after presenting the
dynamics of geometry in terms of best matching.

To conclude the particle dynamics, the strong form of the
Poincar\'e principle does almost everything that one
could ask. It cannot entirely fix the potential term $W$ but
does
require all of its terms to be homogeneous of degree $l^{-2}$ with dimensionless
coefficients, one of which can always be set to unity. If the strong
Poincar\'e principle fails, the most interesting way the weak form can hold 
in the $N$-body problem is if $E=0$. In this case a
one-parameter freedom in the shape-space initial data for given $s$ and $d$ is
associated with the ratio $T_s/T$ in $Q^N$. 

\section{Conformal Geometrodynamics \label{cs}}

Although limited to particle dynamics, the previous section has identified the
two universal elements of shape dynamics: derivation of time from difference
and best matching to obviate the introduction of absolute (nondynamical)
structure. However, nothing can come of nothing. The bedrock on which dynamics
has been derived is the geometrical structure of individual configurations of
the universe. We began with configurations in Euclidean space and removed from
them more and more structure by group quotienting. We left open the question of
how far such quotienting should be taken, noting that nature must decide that.
In this section, we shall see that, with two significant additions, the two
basic principles of shape dynamics can be directly applied to the dynamics of
geometry, or \emph{geometrodynamics}. This will lead to a novel derivation of,
first, general
relativity, \emph{then} special relativity (and gauge theory) and after that to
the remarkable
possibility that gravitational theory introduces a dynamical standard of rest in
a closed universe.

In this connection, let me address a likely worry of the reader,
anticipated in footnote \ref{worry}, about the
fundamental role given to instantaneous configurations of the universe. Does
this not flagrantly contradict the relativity of simultaneity,
which is confirmed by countless experiments? In response, let me mention some
possibly relevant facts.

When Einstein and Minkowski created special relativity, they did not ask how it
is that inertial frames of reference come into existence. They took them as
given. Even when creating
general relativity, Einstein did not directly address the origin of local
inertial frames of reference. Moreover, although he gave a
definition of simultaneity at spatially separated points, he never asked how
temporally separated durations are to be compared. What does it mean to say
that a second today is the same as a second yesterday? Shape dynamics directly
addresses both of these omissions of Einstein, to which may be added his
adoption of length as fundamental, which Weyl questioned in 1918
\cite{Weyl1918, Weyl1997}. Finally, it is a pure historical accident that
Einstein, as he himself said, created general
relativity so early, a decade before quantum
mechanics was discovered. Now it is an architectonic feature of quantum
mechanics that the
Schr\"odinger wave function is defined on configuration space, not (much to
Einstein's dismay) on spacetime.

	This all suggests that instantaneous spatial configurations of the
universe could at the least be considered as the building blocks of
gravitational theory. Indeed, they are in the Hamiltonian dynamical form of
general relativity introduced by Dirac \cite{Dirac1958} and Arnowitt, Deser and
Misner \cite{Arnowitt1962}. However, many relativists regard that formulation as
less fundamental than Einstein's original one. In contrast, I shall argue that
the shape-dynamical approach might be more fundamental and that the
geometrical theory of gravity could have been
found rather naturally using it. I ask the reader to keep an open mind.

\subsection{Superspace and conformal superspace \label{scs}}

Differential geometry begins with the idea of continuity, encapsulated in the
notion of a manifold, the rigorous definition of which takes
much care. I assume that the reader is
familiar with the essentials and also with diffeomorphisms; if not,
\cite{Schutz1980} is an excellent introduction.
To model a closed universe, we need to consider closed manifolds. The simplest
possibility that matches our direct experience of
space is $S^3$, which can be
pictured as the three-dimensional surface of a four-dimensional sphere.

Now suppose that on $S^3$ we define a Riemannian 3-metric $g_{ij}(x)$. 
As a $3\times 3$ symmetric matrix at
each space point, it can always be transformed at 
a given point to diagonal form with $1,1,1$ on the diagonal. Such a metric does
three things. First, it defines  the \emph{length}  $\rm ds$ of the line element
$\rm dx^i$ connecting neighbouring points
of the manifold: $\rm ds=\sqrt{g_{ij}\rm dx^i\rm dx^j}$. This is well known.
However, for shape dynamics it is more important that $g_{ij}(x)$ determines
\emph{angles}. Let two curves at $x$ be tangent to
the line elements $\rm dx^i$ and
$\rm dy^i$ and $\theta$ be the angle between them. Then
\be
\rm{cos}~\theta = {g_{ij}\rm dx^i\rm dy^j\over\sqrt{g_{kl}\rm dx^k\rm
dx^lg_{mn}\rm dy^m\rm dy^n}}.\label{angle}
\ee
The third thing that the metric does (implicitly) is give information about the
coordinates employed to express the metric relations. 

We see here an immediate analogy between a 3-metric and an $N$-body
configuration of particles in Euclidean space. Coordinate information is mixed
up with geometrical information, which itself comes in two different forms:
distances and angles. Let us take this analogy further and introduce
corresponding spaces and structure groups.

\emph{Riem}($S^3$) is the (infinite-dimensional) space of all
suitably continuous Riemannian 3-metrics $g_{ij}$ on $S^3$. Thus, each point in
Riem is a 3-metric. However, many of these 3-metrics express identical distance
relationships on the manifold that are simply expressed by means of different
coordinates, or labels. They can therefore be carried into each other by
three-dimensional diffeomorphisms without these distance relations being
changed. They form
a diffeomorphism equivalence class $\{g_{ij}\}_{diff}$, and the 
3-diffeomorphisms form a structure group that will play a role analogous to the
Euclidean group in particle dynamics. Each such equivalence
class is an orbit of the 3-diffeomorphism group in Riem and is defined as a
\emph{three-geometry}. All such 3-geometries forms \emph{superspace}.
This is a familiar concept in geometrodynamics \cite{Misner1973}. Less
known is \emph{conformal superspace}, which is obtained from superspace by the
further quotienting by conformal transformations:
\be
g_{ij}(x)\rightarrow \phi(x)^4g_{ij}(x),~\phi(x)>0.\label{cons}
\ee
Here, the fourth power of the position-dependent function $\phi$ is chosen for
convenience, since it
makes the transformation of the scalar curvature $R$ simple (in four
dimensions, the corresponding power is 2); the condition $\phi>0$ is
imposed to stop the metric being transformed to the zero matrix. 

The transformations (\ref{cons}) change the distance relations on the manifold
but not the angles between curves. Moreover, distances
are not directly observable. To measure an interval, we must lay a ruler
adjacent to it. If the interval and the ruler
subtend the same angle at our eye, we say that they have the same length.
This is one reason for thinking that angles are more fundamental
than distances; another is that they are dimensionless. We also have the
intuition that shape is more basic than size; we generally
speak of \emph{the}, not an,
equilateral triangle. It is therefore natural to make the combination of the
group of 3-diffeomorphisms and the conformal
transformations (\ref{cons}) the structure group of conformal
geometrodynamics.

Before continuing, I want to mention the subgroup of
the transformations (\ref{cons}) that simply multiply the 3-metric by a 
constant
$C$:
\be
g_{ij}(x)\rightarrow Cg_{ij}(x),~C>0.\label{constcons}
\ee

One can say that the transformations (\ref{constcons}) either
`change
the size of the universe' or change the unit of distance. Like similarity
transformations, they leave all length ratios unchanged, and are
conceptually distinct from the general transformations (\ref{cons}), which
change the ratios of the geodesic lengths $d(a,b)$ and $d(c,d)$ between point
pairs $a,b$ and $c,d$. As a result, general conformal transformations open up a
vastly richer field for study than similarity transformations. Another
subgroup consists of the \emph{volume-preserving
conformal transformations} (\ref{cons}). They leave the total
volume
$V=\int\sqrt g\rm d^3x$ of the universe unchanged. 
We shall see that these transformations play an
important role in cosmology. The seemingly minor restriction of
the transformations (\ref{cons}) to be volume preserving is the mysterious
last vestige of absolute space that I mentioned in the introduction.

The idea of geometrodynamics is nearly 150 years old. Clifford, the
translator of Riemann's 1854 paper on the foundations of geometry,
conjectured in 1870 that material bodies in motion
might be nothing more than regions of empty but
differently curved three-dimensional space moving relative to each other
\cite{Misner1973}, p. 1202. This
idea is realized in Einstein's general relativity in the vacuum (matter-free)
case in the geometrodynamic interpretation advocated by Wheeler
\cite{Misner1973}. I shall
briefly describe his superspace-based picture, before taking it further
to conformal superspace.

Consider a matter-free spacetime that is globally hyperbolic. This means
that one can slice it by nowhere intersecting spacelike hypersurfaces identified
by a monotonic time label $t$ (Fig.~7). Each hypersurface carries a 3-geometry,
which 
can be represented by many different 3-metrics $g_{ij}$. At any point $x$ on one
hypersurface labelled by $t$ one can move in spacetime orthogonally to the
$t+\delta t$ hypersurface, reaching it after the proper time
$\delta\tau=N\delta t$, where $N$ is called the \emph{lapse}. If the time
labelling is changed, $N$ is rescaled in such a way that $N\delta t$ is
invariant. In general, the coordinates on successive 3-geometries will be
chosen arbitrarily, so that the point with coordinate $x$ on hypersurface
$t+\delta t$ will not lie at the point at which the normal erected at point $x$
on hypersurface $t$ pierces hypersurface $t+\delta t$. There will be a lateral
displacement of magnitude $\delta x^i=N^i\delta t$. The vector $N^i$ is called
the \emph{shift}. The lapse and shift encode the $g_{00}$
and $g_{0i}$ components respectively of the 4-metric: $g_{00}=N_iN^i-N^2,
g_{0i}=N_i$.

\begin{figure}
 \begin{center}
 \includegraphics[width=11cm]{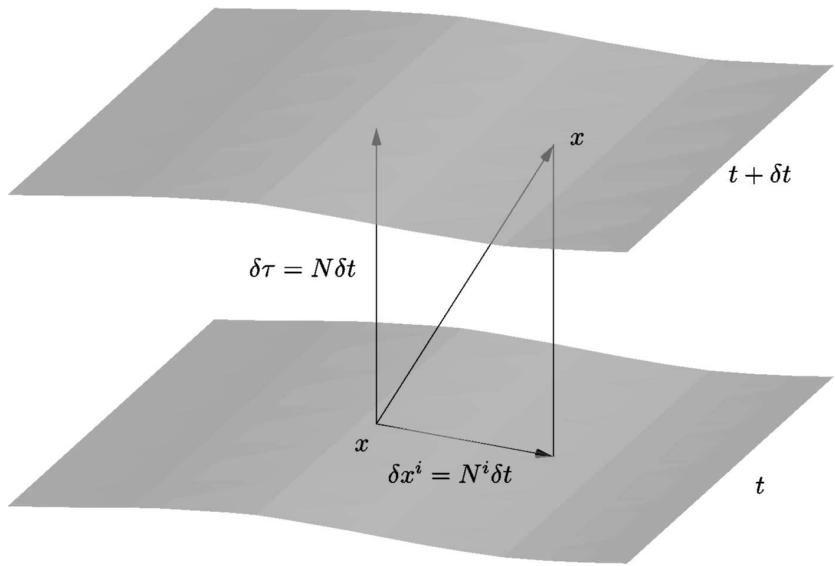}

\caption{The $3+1$ decomposition of spacetime as explained in the text.}

 \end{center}

\end{figure}

Each 3-metric $g_{ij}$ on the successive hypersurfaces is a point in Riem, and
the one-parameter family of successive $g_{ij}$'s is represented as a curve in
Riem
parametrized by $t$. This is just one representation of the spacetime.
First, one can change the time label freely on the curve (respecting
monotonicity). This leaves the curve in Riem unchanged and merely changes its
parametrization. Second, by changing
the spatial coordinates on each hypersurface one can change the
successive 3-metrics and move the curve around to a considerable degree in
Riem. However, each of these curves corresponds to one and the same curve in
superspace. But, third, one and the same
spacetime can be sliced in many different ways because
the definition of simultaneity in general
relativity is to a high degree arbitrary (Fig.~8). Thus, an
infinity of curves in superspace, and an even greater
infinity of curves in Riem, represent the same spacetime. In addition,
they can all carry infinitely many different parametrizations by time labels.
This huge freedom corresponds to the possibility of making arbitrary
four-dimensional coordinate transformations, or equivalently
4-diffeomorphisms, on spacetime. 

\begin{figure}
 \begin{center}
 \includegraphics[width=11cm]{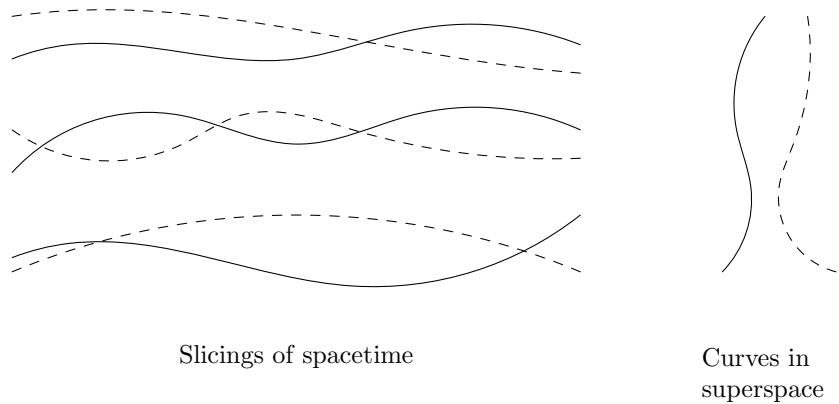}

\caption{Because there is no distinguished definition of simultaneity in
general relativity, a spacetime can be sliced in many different
ways. This slicing, or foliation, freedom leads to many different
representations of the spacetime by curves in superspace. Two slicings and
corresponding curves in superspace are shown.}

 \end{center}

\end{figure}

As long as one insists on the equal status of all different slicings by
spacelike hypersurfaces -- on \emph{slicing} or \emph{foliation} invariance -- 
it is not possible to represent the evolution of 3-geometry by a unique curve
in a geometrical configuration space. This is the widely accepted view of
virtually all relativists. Shape dynamics questions this.
I shall now sketch the argument.

Purely geometrically, distinguished foliations in
spacetime \emph{do exist}. The
flat \emph{intrinsic} (two-dimensional) geometry of a sheet of paper is
unchanged when
it is rolled into a tube and acquires
\emph{extrinsic curvature}. By analogy, just as a 3-metric $g_{ij}$
describes intrinsic geometry, a second fundamental form, also a $3\times
3$ symmetric tensor $K^{ij}$, describes extrinsic curvature. Its trace
$K=g_{ig}K^{ij}$ is the
\emph{mean extrinsic curvature}. A \emph{constant-mean-curvature} (CMC)
hypersurface is one embedded in spacetime in
such a way that $K$ is everywhere constant. In three-dimensional
Euclidean space
two-dimensional soap bubbles have CMC surfaces. Such surfaces are extremal and
are therefore associated with `good' mathematics. At least geometrically, they
are clearly distinguished. 

A complete understanding of the possibilities for slicing a spatially closed
vacuum Einsteinian spacetime, i.e., one that satisfies Einstein's field
equations $G_{\mu\nu}=0$, by CMC hypersurfaces does not yet
exist.\footnote{There certainly exist spacetimes that satisfy Einstein's field
equations and do not admit CMC foliation. However, shape dynamics does not have
to yield all solutions allowed by general
relativity but only those relevant for the description
of the universe. The ability to reproduce nature, not general
relativity, is what counts.}
However, as
we shall see, there exists a very effective and reliable way to generate
`patches' of CMC-foliated Einsteinian spacetimes. In such a patch CMC-foliated
spacetime exists in an open neighbourhood either side of
some CMC hypersurface labelled by $t=0$. A noteworthy
property of CMC foliations is that $K$, which is necessarily a spatial
constant on each hypersurface, must change
monotonically in the spatially closed case. 
Moreover, $K$ measures the rate of change of the spatial volume $V=\int\sqrt
g\,\rm d^3x$ in unit proper time.\footnote{The value of $K$ is only
defined up to its sign.} In both these respects, $K$ is closely
analogous to the quantity $D$ (\ref{D}) in particle mechanics. We recall that
it is the rate of increase of the moment of inertia $I$, which, like $V$,
characterizes the size of the universe. 

Let us now suppose that we do have a vacuum Einsteinian spacetime that is CMC
foliated either in its entirety or in some patch. On each leaf (slice) of the
foliation there will be some 3-geometry and a uniquely determined conformal
3-geometry, i.e., that part of the 3-geometry that relates only to angle
measurements. We can take the successive conformal 3-geometries and plot them
as a curve in conformal superspace (CS). Having done this, we could change the
slicing in the spacetime, obtaining a different curve of 3-geometries in
superspace. They too would have associated conformal 3-geometries, and each
different curve of 3-geometries in superspace would generate a different curve
of conformal 3-geometries in CS. According to the standard
interpretation of
general
relativity, all these different curves in superspace and in CS are to
be regarded as physically equivalent. I believe there are grounds to at least
question this.

If we go back to Clifford's original inspiration, note that only angles are
observable, and insist on either the strong or weak form of the Poincar\'e
principle, we are led naturally to the desire to create a dynamical theory of
conformal geometry in which either a point and a direction in CS or a point and
a tangent vector in CS suffice to determine a unique evolution in CS. This will
be exactly analogous to the aim of particle shape dynamics and will be
implemented in the following subsections. What makes this
shape-dynamic approach interesting is that the successions of conformal
3-geometries generated in the weak case correspond exactly to the
successions of conformal 3-geometries obtained on CMC foliated Einsteinian
spacetimes. Moreover, the best matching by which the dynamic curves in CS are
obtained simultaneously generates the complete spacetime as the
distinguished representation of the conformal dynamics. Once this spacetime
has been generated in the CMC foliation, one can go over to an arbitrary
foliation within it and recover all of the familiar results of general
relativity. Three distinct ingredients create conformal
dynamics. I shall present them one by one.

\subsection{The elimination of time}

It is easy (Sec.~\ref{jacobi}) to remove time from the kinematics
of particle 
dynamics and recover it as a distinguished parameter from geodesic dynamics. It
will help now to look at the structure of the canonical momenta
in relational particle dynamics. Given a Lagrangian $L(q_a,q_a')$ that
depends
on dynamical variables $q_a$ and their
velocities $\,q_a'=\rm
dq_a/\rm d\lambda$, the canonical momentum of $q_a$ is
 $p^a:=\partial L/\partial q_a'$. For the best-matched action (\ref{bma}), 
\be
\bf p^a:={\partial L^{bm}\over\partial \bf{x}_a'}=\sqrt{W\over
T_{bm}}m_a{\rm
d\bf x_a^{bm}\over\rm d\lambda},~L^{bm}=\sqrt{WT^{bm}}.\label{cm}
\ee
The distinguished time label $t$ is obtained by choosing 
$\lambda$ such that always $W=T_{bm}$ so that the cofactor of $m_a{\rm
d\bf x_a^{bm}/\rm d\lambda}$ is unity. The definition is
holistic for two reasons. First, the ${\rm d\bf
x_a^{bm}/\rm d\lambda}$ are obtained by global best matching and
are therefore determined by all  the changes of the relative
separations of the particles. Second, the denominator of the factor
$\sqrt{W/T_{bm}}$ is a sum over the displacements of all the particles in the
universe. This is seen explicitly in the expression (\ref{newt}). The $\bf{p}^a
$ have a further important property: they are reparametrization invariant.
If one rescales $\lambda, \lambda\rightarrow \bar\lambda(\lambda)$, the
velocities
$q_a'$ scale, but because velocities occur
linearly in the numerator and denominator of (\ref{cm}) there is no change in
$\bf p^a$, which is in essence is a \emph{direction}. (It is in fact a
direction cosine wrt to the conformal metric obtained by multiplying the
kinetic metric by $W$.)

However, one could take the view that, at any instant, one should obtain a
\emph{local} measure of time derived from purely local differences. This would
still yield an holistic notion of time if the local differences were obtained
by best matching. However, in the case of particle dynamics, a local derived
time of this kind cannot be obtained for the simple reason that particles are,
by definition, structureless. The situation is quite different in field
dynamics because fields have several components at each space point. This opens
up the possibility of a local measure of time, as I shall now show (deferring
the conceptually distinct issue of best matching until later).

Let
the action on Riem, the configuration space in which calculations are of
necessity made, have the form
\be
I=\int\rm d\lambda L,~ L=\int\rm d^3x\sqrt{gWT},\label{sqrt1}
\ee
where $g=\sqrt{\rm{det}~g_{ij}}$ is introduced
explicitly to make the integrand a tensor density, the scalar $W$ is a
local functional of $g_{ij}$ (that is, it depends on $g_{ij}$ and its spatial
derivatives up to some finite order, which will in fact be the second), and $T$
depends quadratically on the metric velocities $g_{ij}':=\rm dg_{ij}/\rm
d\lambda$ and also quadratically on $g^{ij}$. It will actually have the form
\be
T=G_A^{\,ijkl}g_{ij}'g_{kl}',
~G_A^{\,ijkl}=g^{ik}g^{jl}+Ag^{ij}g^{kl}.\label{smA}
\ee
Here, $G_A^{\,ijkl}$, in which $A$ is an as yet arbitrary constant, is
a supermetric (cf (\ref{metric})) and appears because one needs to construct
from the
velocities 
$g_{ij}'$ a quantity that is a scalar under 3-diffeomorphisms. Because 
$g_{ij}'$ is, like the 3-metric, a symmetric tensor, there are only
two
independent scalars that one can form from it by contraction using the inverse
metric.\footnote{The most general supermetric formed from $g^{ij}$ and acting on
a general tensor has three terms. For $A=-1$, we obtain the DeWitt supermetric,
which will appear later. In principle, one could also consider supermetrics
formed with spatial derivatives of $g^{ij}$, but these would lead to very
complicated theories. As Einstein always recommended, it is advisable to look
first for the
simplest nontrivial realizations of an idea.}

The key thing about (\ref{sqrt1}) is that one first forms a quantity
quadratic in the velocities at each space point, takes the square root at each
space point, and only then integrates over space. This is a 
\emph{local square root} and can be justified as follows. First, a square root
must be introduced in \emph{some} way to create a theory without any
external time variable. Next, there are two ways in which this can be done. The
first is by direct analogy with Jacobi's principle (\ref{metric1}) or
(\ref{metric2}), and would lead to an action with `global' square roots of the
form
\be
I=\int\rm d\lambda\sqrt{\int\rm d^3x\sqrt{g}W}\sqrt{\int\rm
d^3x\sqrt{g}T}.\label{sqrt2}
\ee
Besides being a direct generalization, the action (\ref{sqrt2}) is on the face
of it mathematically more correct than (\ref{sqrt1}) since it defines a proper
metric on Riem, which is not the case if the square root is
taken, as in (\ref{sqrt1}), before the integration over space. Nevertheless, it
turns out that an action of the form (\ref{sqrt1}) does lead to a consistent
theory. This will be shown below, but we can already see that in such a case we
obtain a theory with a local emergent time. For this, we
merely need to calculate the form of the canonical momenta of the 3-metric
$g_{ij}$ that follow from (\ref{sqrt1}):
\be
p^{ij}:={\partial L\over\partial g_{ij}'}=\sqrt{W\over
T}G^{ijkl}g_{kl}'.\label{cmg}
\ee
The similarity of the $p^{ij}$ to the particle canonical momenta
$\bf{p}^a$ (\ref{cm}) is obvious. First, under
$\lambda\rightarrow\bar\lambda(\lambda),$
the momenta $p^{ij}$ are, like $\bf{x}^a$, unchanged. Second, the complex of
bare
velocities $G^{ijkl}g_{kl}'$ is multiplied by the Jacobi-type factor
$\sqrt{W/T}$. However, the key difference is that this factor is no longer a
global but a position-dependent local quantity. I will not go into further
details yet except to say that when the theory
is fully worked out it leads to the appearance of a local increment of proper
time given by $\delta\tau=N\delta\lambda$, where $N=\sqrt{T/4R}$ can be
identified with the lapse in general relativity.

Whereas the elimination of time in Jacobi's principle and for an action
like (\ref{sqrt2}) with global square roots is, at the classical level at
least,\footnote{In quantum mechanics, the effect is dramatic, since the
quantization of Jacobi's principle leads to a time-independent, and
 not time-dependent Schr\"odinger equation. This is one aspect of the famous
`problem of time' in canonical quantum gravity \cite{Kuchar1992,
Isham1993,Barbour1999a}.} a trivial matter with no impact on the best matching
(and vice versa), the
elimination of external time by the local square root has a huge effect and its
consequences become intimately interconnected with those of the best matching.
Perhaps the most important effect is that it drastically reduces the number
of \emph{consistent} actions that one can construct. This was first recognized
by my collaborator Niall \'O Murchadha, and its consequences were explored
in \cite{Barbour2002}, about which I shall say something after the description
of geometrodynamic best matching.

\subsection{Geometrodynamic best matching}

The basic idea of geometrodynamic best matching is exactly as for
particles but leads to a vastly richer theory because a 3-geometry, either
Riemannian or conformal, is
infinitely more structured than a configuration of particles in Euclidean space.
However, the core idea is the same: to
`minimize the incongruence' of two intrinsically distinct
configurations. This is done by using the spatial structure
groups of the configurations to bring one configuration into the position in
which it most closely overlaps the other.

Let us first consider 3-diffeomorphisms. If we make an
infinitesimal coordinate transformation on a given 3-metric $g_{ij}(x)$,
obtaining new functions of new coordinates, $g_{ij}(x)\rightarrow \bar
g_{ij}(\bar x)$, and then consider $\bar
g_{ij}(\bar x)$ at the old $x$ values, the resulting 3-metric $\bar
g_{ij}(x)$ is what one obtains by a 3-diffeomorphism generated
by some 3-vector field $\xi^i(x)$: $\bar g_{ij}(x)=g_{ij}(x)+\xi_{(i;j)}$
(the semi-colon denotes the covariant
derivative wrt to $g_{ij}$ and the round parentheses symmetrization). The
 two 3-metrics $g_{ij}(x)$ and $\bar g_{ij}(x)$ are diffeomorphically related
representations of one and the same 3-geometry. This is analogous to
changing the Cartesian coordinates of a particle configuration.

Now suppose that $g_{ij}(x)+\delta g_{ij}(x)$ represents a 3-geometry genuinely
distinct from $g_{ij}(x)$, i.e., $\delta g_{ij}$ cannot
be represented in the form $\xi_{(i;j)}$, which would indicate a spurious
diffeomorphically-induced change. The difficulty that we now face
is that, because we are considering intrinsically different 3-geometries, mere
identity of the coordinate values $x_i$ does not mean that they specify `the
same point' in the two different 3-geometries. In fact, the problem is nothing
to do with coordinates. Given an apple and a pear, there does not appear to be
any way to establish a 1-to-1 pairing of all the points on the apple's
surface with all those on the pear's.
However, best matching does just that if the compared objects differ only
infinitesimally. To apply the technique rigourously, one must use
rates of change rather than finite differences.

Mathematically, we can always specify a 3-metric $g_{ij}(x)$
and its velocity $g_{ij}'=\rm dg_{ij}/\rm d\lambda$. The problem is that 
$g_{ij}'=\rm dg_{ij}/\rm d\lambda$ mixes information about the intrinsic
change
of the described 3-geometry with
arbitrary information about the way in which the coordinates are laid down as
the 3-geometry changes. There is an equivalence class of velocities
$\{g_{ij}'-\xi_{(i;j)}'\}$ that all represent the same intrinsic
change.
The
task of best matching is to select a unique one among them that can be said to
measure the true change. 

We note first that there is no objection to fixing coordinates on the original
3-geometry, giving $g_{ij}(x)$, just as we chose an initial Cartesian
representation for the particle configurations. To fix the way the coordinates
are then laid down, we consider the effect of $\lambda$-dependent
diffeomorphisms on (\ref{sqrt1}). It becomes
\be
I=\int\rm d\lambda L,~ L=\int\rm
d^3x\sqrt{gWT},~T=G_A^{\,ijkl}(g_{ij}'-\xi_{(i;j)}')(g_{kl}'-\xi_
{(k;l)}').\label {sqrt3}
\ee

The possibility of constructing consistent geometrodynamical theories is
considered in \cite{Barbour2002},\footnote{Some of the conclusions reached in
\cite{Barbour2002} are too
strong, being based on tacit simplicity assumptions that Anderson identified
\cite{Anderson2007, Anderson2007a}. I shall report here the most interesting
results that are obtained when the
suitable caveats are made.} to which I refer the reader for details,
since I only wish to indicate what the results are.

The basic theoretical structure obtained in geometrodynamics
is broadly the same as in particle dynamics. One obtains constraints and
conditions under which they propagate consistently. These conditions
strongly restrict the set of consistent theories. I shall first identify the
constraints and then indicate how they act as `theory selectors'.

First, there are constraints
because of the local square root in (\ref{sqrt3}).
Before giving them, I need to draw attention to a similar constraint, or
rather identity, in the particle model. It follows from the form
(\ref{cm}) of the canonical momenta $\bf p^a$ that
\be
\sum_a{\bf p^a\cdot\bf p^a\over 2m_a}\equiv W.\label{cmc}
\ee
This is a \emph{square-root identity}, since it
follows directly from the square root in the Lagrangian and means
that the $\bf p^a$ are in essence direction cosines. In the Hamiltonian
formalism, (\ref{cmc}) becomes a constraint and is a
single global relation. In contrast, the
geometrodynamic action contains an infinity of square roots, one at each space
point. Correspondingly, the canonical momenta satisfy \emph{infinitely} many
identities (or Hamiltonian constraints):
\be
p_{ij}p^{ij}-{2A\over{3A-1}}p^2\equiv gW,~p=g_{ij}p^{ij}.\label{qc}
\ee

Second, constraints arise through the best matching wrt diffeomorphisms. This
is implemented by variation of (\ref{sqrt3}) with respect to $\xi_i^{\,'}$,
treated
as a Lagrange multiplier. This also leads to a constraint at each space point:
\be
p^{ij}_{~;j}=0.\label{momc}
\ee
These linear constraints are closely analogous to the linear constraints
(\ref{P}) and (\ref{L}) in particle dynamics. For the form
(\ref{sqrt3}) of the action, they propagate automatically and do not lead to
restrictions. This is because the action (\ref{sqrt1}) was chosen in
advance in a form invariant under
$\lambda$-independent 3-diffeomorphisms, which in turn ensured that
(\ref{sqrt3}) is invariant under $\lambda$-dependent 3-diffeomorphisms. Had we
chosen a
more general functional of $g_{ij}$ and its spatial and $\lambda$
derivatives, propagation of the constraints (\ref{momc})
would have forced us to specialize the general form to (\ref{sqrt3}). This is
another manifestation of the power of combining the structure group of the
3-metrics (the 3-diffeomorphism group) with the Poincar\'e requirement.

There is no analogous control over the quadratic constaints (\ref{qc}) that
arise from the local square root in (\ref{sqrt1}) and (\ref{sqrt3}). As is
shown in \cite{Barbour2002}, the only actions that consistently propagate both
the
quadratic and linear constraints has the form
\be
I_{BSW}=\int\rm d\lambda\int\rm d^3x\sqrt{(\Lambda+dR)T_{A=-1}},\label{BSW}
\ee
where the subscript $A=-1$ of $T$ indicates that the undetermined coefficient in
the supermetric is forced to take the DeWitt value. More impressive is the
drastic restriction on the possible form of the potential term $W$, which is
restricted to be $\Lambda + dR, d=0~\rm{or}~\pm 1.$ The action (\ref{BSW}) is in
fact the
Baierlein--Sharp--Wheeler action \cite{Baierlein1962}, which is dynamically
equivalent to the
Einstein--Hilbert action for globally hyperbolic spacetimes. The only freedom
is in the choice of the constant $\Lambda$, which corresponds to the
cosmological constant, and the three options for $d$. The case $d=0$ yields
so-called strong gravity and is analogous to pure inertial motion for
particles. The case $d$ corresponds to a Lorentzian spacetime and hence to the
standard form of general relativity, while $d=-1$ gives Euclidean general
relativity.

When translated into spacetime terms,\footnote{The 
`construction of spacetime' will be described later.} the constraints (\ref{qc})
and (\ref{momc}) are respectively the $00$ and $0i, i=1,2,3,$ Einstein field
equations $G_{\mu\nu}=0, \mu,\nu=0,1,2,3.$ Whereas the
particle dynamics associated with the global Euclidean group
leads to global relations, implementation of the Poincar\'e principle in
geometrodynamics by the local elimination of time and best matching wrt local
3-diffeomorphisms leads to local constraints, the propagation of which directly
determines the simplest nontrivial realization of the whole idea: general
relativity. Of course, the immense power of local symmetry requirements was one
of the great discoveries of 20th-century physics. It first became apparent
with Einstein's creation of general
relativity. If shape dynamics has value, it is not so much
in the locality of the symmetries as in their choice and in the treatment of
time. I shall compare the shape-dynamic approach with Einstein's at the end of
the paper. Here I want to continue with the results of
\cite{Barbour2002}.

So far, we have considered pure geometrodynamics. The assumption that the
structure of spacetime always reduces locally to the Minkowski-space form of
special relativity (a key element in Einstein's approach) played no role in the
derivation. The manner in which special relativity arises in
\cite{Barbour2002} is striking. In field theory, the essence
of special relativity is a \emph{universal light cone}: all fields must have
the same limiting signal propagation velocity. Now vacuum general
relativity has a `light
cone'. What happens if we attempt, as the simplest possibility, to couple a
scalar field $\varphi$ to vacuum geometrodynamics described by the action
(\ref{sqrt3})? 

The propagation speed of such a field, with action containing
the field velocities $\rm d\varphi/\rm d\lambda$ and first spatial derivatives
$\partial_i\varphi$ quadratically, is determined by a single coefficient $C$,
which
fixes the ratio of the contributions of  $\rm d\varphi/\rm d\lambda$ and 
$\partial_i\varphi$ to the action. When the scalar field is added to the action
(for details see \cite{Barbour2002}), the constraints (\ref{qc}) and
(\ref{momc}) acquire additional terms, and one must verify that the modified
constraints are propagated by the equations of motion. It is shown in
\cite{Barbour2002} that propagation of the modified quadratic constraint fixes
the coefficient $C$ to be exactly such that it shares the geometrodynamic light
cone. Otherwise, the scalar field can have a term in its action corresponding
to a mass and other self-interactions. 

The effect of attempting to couple a single 3-vector field $\bf A$ to the
geometry is even more remarkable. In this case, there are three possible 
terms that can be formed from the first spatial (necessarily covariant)
derivatives of $\bf A$. Each may enter in principle with an arbitrary
coefficient. The requirement that the modified quadratic constraint propagate
not only fixes all three coefficients in such a way that the 3-vector field has
the same light cone as the geometry but also imposes the requirement that the
canonical momenta $\bf P$ of $\bf A$ satisfy the constraint $\rm{div}~\bf P=0$.
In fact, the resulting field is none other than the Maxwell field interacting
with gravity. The constraint $\rm{div}~\bf P=0$ is the famous Gauss constraint.
This can be taken further \cite{Anderson2002}. If one attempts to construct a
theory of several
3-vector fields that interact with gravity and with each other, they have to be
Yang--Mills gauge fields. Unlike the scalar field, all the gauge fields must be
massless.

To conclude this subsection, let us indulge in some `what-might-have-been'
history.  Clifford's `dream' of explaining all motion
and matter in terms of dynamical Riemannian 3-geometry was in essence a
proposal for a new ontology of the world. The history of science shows that
new, reasonably clearly defined ontologies almost always precede major advances.
A good example is Descartes's formulation of the mechanical world view; it led
within a few decades to Newton's dynamics (\cite{Barbour2001}, Chaps. 8--10).
Clifford died tragically young; he could have lived to interact with both Mach
and Poincar\'e. Between them, they had the ideas and ability needed to create a
relational theory of dynamical geometry (and other fields) along the lines
described above. In this way, well before 1905, they could have discovered,
first, general relativity in the form of the Baierlein--Sharp--Wheeler action
(\ref{BSW}), next special relativity through a universal light cone, and even,
third, gauge theory. All of this could have happened as part of a programme to
realize Clifford's original inspiration in the simplest nontrivial way.

I want to emphasize the role that the concept of time would have
played in such a scenario. In 1905, Einstein transformed physics by insisting
that the description of motion has no meaning ``unless we are quite clear as to
what we understand by `time' '' \cite{Einstein1905}. He had in mind the
problem of defining simultaneity at spatially separated points. Resolution of
this
issue in 1905 was perhaps the single most important thing that then led on to
general relativity. However, in 1898, Poincar\'e \cite{Poincar'e1898,
Poincar'e1904} had noted the existence of \emph{two} fundamental problems
related to time: the definition of simultaneity and the
older problem of defining duration: What does it mean to say that a second
today is the same as a second tomorrow? Even earlier, in 1883, Mach had said:
``It is utterly beyond our power
to measure the changes of things by time. Quite the contrary, time is an
abstraction at which we arrive by means of the changes of things.'' Both Mach
and Poincar\'e had clearly recognized the need for a theory of duration along
the lines of Sec.~\ref{jacobi}.\footnote{Despite a careful search through his
papers, I have been unable to find any evidence that Einstein ever seriously
considered the definition of duration. As we saw in Sec.~\ref{jacobi}, this is
intimately related to the theory of clocks, which Einstein
did grant had not been properly included in general relativity. He called the
omission a `sin' \cite{Einstein1949}.} 

There is now an intriguing fact. The structure of dynamics so far
presented in this paper has been based on two things: best matching and a theory
of duration. Both were initially realized globally, after which a local
treatment was introduced. Moreover, entirely different schemes were used to
achieve the desired aims of a relational treatment of displacement and
of duration
(best matching and a square root in the action respectively). Remarkably,
Einstein's theory of simultaneity appeared as a \emph{consequence} of these
relational inputs. In line with my
comments at the end of Sec.~\ref{jacobi}, I believe that the 
concept of duration as a measure of difference is more fundamental than the
definition of simultaneity, so it is reassuring that Einstein's well confirmed
results can be recovered starting from what may be deeper foundations. In this
connection, there is another factor to consider. In the standard representation
of general relativity, spacetime is a four-dimensional block. One is not
supposed to think that the Riemannian 3-geometry on the leaves of a 3+1
foliation is more fundamental than the lapse and shift, which tell one how the
3-geometries on the leaves `fit together' (Fig.~7). The lapse is particularly
important: 
it tells you the orthogonal separation (in spacetime) between the 3-geometries
that are the leaves of a 3+1 foliation. However, the $G^{00}$ Einstein field
equation enables one to solve algebraically for the lapse in terms of the other
variables. It
is precisely this step that led Baierlein, Sharp and Wheeler to the BSW action
(\ref{BSW}). It contains no lapse, but, as we have seen, is exactly the kind of
action that one would write down into to implement (locally) Mach's requirement
that time (duration) be derived from differences. Thus, there is an exactly
right theory of duration at the heart of general relativity, but it is hidden
in the standard representation.

However, this is not the end of the story. Quite apart from the implications of
the two aspects of time -- duration and simultaneity -- for the quantum theory
of the universe, there is also what Weyl \cite{Weyl1952} called the ``disturbing
question of length'': Why does nature seem to violate the principle that size
should be relative? We shall now see that a possible answer to this question
may add yet another twist to the theory of time.

\subsection{Conformal best matching}

In best matching wrt 3-diffeomorphisms, we are in effect looking at all
possible ways in which all points on one 3-geometry can be mapped bijectively
to the points of an intrinsically different 3-geometry and selecting the
bijection that extremalizes\footnote{We have to extremalize rather than minimize
because the DeWitt supermetric ($G^{ijkl}_{A=-1}$ in (\ref{smA})) is indefinite.
Einsteinian gravity is unique
among all known physical fields in that its kinetic energy is not positive
definite. The
part associated with expansion of space -- the second term in
(\ref{sqrt3}) -- enters with the opposite sign to the part associated with the
change of the conformal part of the 3-metric, i.e., its shape.} the quantity
chosen to measure the incongruence of the two. So far, we have not
considered changing the local scale factor of the 3-metrics in
accordance with the conformal transformations (\ref{cons}). But given that
only angles
are directly observable, we have good grounds for supposing that lengths should
not occur as genuine dynamical degrees of freedom in the dynamics of geometry.
If we best match wrt conformal transformations, only the
angle-determining part of 3-metrics can play a dynamical role. Moreover, we
have
already noted (Sec.~\ref{scs}) the possibility that we might wish to best match
only wrt
volume-preserving conformal transformations.

At this point, it is helpful to recall the geometrical description of best
matching in the particle model. It relies on a supermetric on the `large'
configuration space $Q^{N}$, which is foliated by the
orbits of whatever group one is considering. Each orbit represents the
intrinsic physical configuration of the system. Hitherto the supermetric
chosen on the
`large' space ($Q^{N}$ or Riem) has been equivariant, so that the
orthogonal separation $\rm ds$ between neighbouring orbits is the same at all
points on the orbits. This made it possible to calculate the orthogonal $\rm
ds$ anywhere between the orbits and, knowing that the same value would always
be obtained, project any such $\rm ds$ down to the physical quotient space.
This met the key aim -- to define a metric on the physical space.

Now there is in principle a different way in which this aim can be met. It
arises if the orthogonal separation between the orbits is not constant
\emph{but has a unique extremum at some point between any two considered
orbits}.\footnote{To the best of my knowledge, this possibility (which certainly
does not occur in gauge theory) was first considered by \'O Murchadha, who
suggested it as a way to implement conformal best matching in
\cite{Barbour1999}.} This unique extremal value can then be taken to
define the required
distance on the physical quotient space. I shall now indicate how this
possibility can be implemented. Since the equations become rather complicated, I
shall not attempt to give them in detail but merely outline what happens. 

We start with the BSW action (\ref{BSW}), since our choices have already been
restricted to it by the local square root and the diffeomorphism best matching
(neither of which we wish to sacrifice, though we will set $\Lambda=0$ for
simplicity). As just anticipated, we immediately encounter a significant
difference from the
best-matching wrt to 3-diffeomorphisms, for which we noted that (\ref{sqrt1}) is
invariant under $\lambda$-independent diffeomorphisms. In the language of gauge
theory, (\ref{sqrt1}) has a global (wrt $\lambda$) symmetry that is subsequently
gauged by
replacing the bare velocity $g_{ij}'$ by the corrected velocity
$g_{ij}'-\xi_{(i;j)}'$. It is the global symmetry which ensures that the
inter-orbit
separation in Riem is everywhere constant (equivariance). In contrast to the
invariance of (\ref{sqrt1}) under
$\lambda$-independent diffeomorphisms, there is no invariance of (\ref{sqrt1})
under $\lambda$-independent conformal transformations of the form (\ref{cons}).
The kinetic term by itself is invariant, but $\sqrt{gR}$ is not. Indeed,
\be
\sqrt {gR}\rightarrow \sqrt g\phi^4\sqrt{R-8{\nabla^2\phi\over\phi}}.  
\ee

It should however be stressed that when (\ref{sqrt1}) is `conformalized' in
accordance with (\ref{cons}) the resulting action is invariant under the
combined gauge-type transformation
\be
g_{ij}\rightarrow \omega^4g_{ij},~\phi\rightarrow {\phi\over\omega},\label{cgi}
\ee
where $\omega=\omega(x,\lambda)$ is an arbitrary function. This exactly matches
the invariance of (\ref{sqrt3}) under 3-diffeomorphisms that arises because the
transformation of $g_{ij}'$ is offset by a compensating transformation of
the
best-matching correction $\xi_{(i;j)}'$. The only difference is that under
the
diffeomorphisms the velocities alone are transformed because of the prior
choice of an action that is invariant under $\lambda$-independent
transformations, whereas (\ref{cgi}) generates transformations of both the
dynamical variables and their velocities.

We now note that if we best match (\ref{sqrt3}) wrt unrestricted conformal
transformations, we run into a problem since we can make the action ever
smaller by taking the value of $\phi$ ever smaller. Thus, we have no
chance of finding an extremum of the action. There are two ways in which this
difficulty can be resolved. The first mimics what we did in particle dynamics
in order to implement the strong Poincar\'e principle on shape space, namely
use a Lagrangian that overall has length dimension zero. 

In the particle model we did this by dividing the kinetic metric $\rm ds$ by
$\sqrt I_{cms}$, where
$I_{cms}$ is the cms moment of inertia. The analog of $I_{cms}$ in
geometrodynamics is $V$,
the total volume of the universe, and division of the Lagrangian in
(\ref{sqrt1}) by $V^{2\over 3}$ achieves the desired result. This route is
explored in \cite{Anderson2003}. It leads to a theory on conformal superspace
that satisfies
the strong Poincar\'e principle and is very similar to general
relativity, except for an
epoch-dependent emergent cosmological constant. This has the effect of enforcing
$V=\rm{constant}$, with the consequence that the theory is incapable of
explaining the diverse cosmological phenomena that are all so well explained by
the theory of
the expanding universe. The theory is not viable. 

An alternative is to satisfy the weak Poincar\'e principle by restricting the
conformal transformations (\ref{cons}) to be such that they leave
the total volume unchanged. At the end of Sec.~\ref{scs}, I briefly
described the consequences. Let me now give 
more details; for the full theory, see \cite{Anderson2005}. The physical space
is initially chosen to be
conformal superspace (CS), to which the space V of possible volumes $V$ of the
universe is adjoined, giving the space CS+V. One obtains a theory that in
principle yields a unique curve between any two points in CS+V. These two
points are specified by giving two conformal geometries $c_1$ and $c_2$, i.e.,
two points in CS, and associated volumes $V_1$ and $V_2$. However, there are two
caveats. First, one cannot guarantee monotonicity of
$V$. This difficulty can be avoided by passing from $V$ to
its canonically conjugate variable; in spacetime terms, this turns out to
be $K$, the constant mean curvature of CMC hypersurfaces. Second, both $V$ and
$K$ have dimensions and as such have no direct physical significance. Only the
curves projected from CS+V to CS correspond to objective reality. In fact,
a two-parameter family of curves in CS+V projects to a single-parameter
family of curves in CS labelled by the dimensionless values of $V_2/V_1$ or,
better, the monotonic $K_2/K_1$.\footnote{This fact escaped notice in
\cite{Anderson2005}. Its detection led to \cite{Barbour2010a}, which shows
that a point and a tangent vector in CS are sufficient to determine the
evolution in CS. In turn, this means that the evolution is determined by
exactly four local Hamiltonian shape degrees of freedom per space point. The
paper \cite{Anderson2005} was written in the mistaken belief that one extra
global degree of freedom, the value of $V$, also plays a true dynamical
role.\label{mb}}

A comparison with the standard variational principle for the $N$-body problem is
here helpful. In it one specifies initial and final configurations in $Q^{N}$,
i.e.,
$2\times 3N$ numbers, together with a time \emph{difference} $t_2-t_1$. Thus,
the variational problem is defined by $6N+1$ numbers. However, the initial
value problem requires only $6N$ numbers: a point in $Q^{N}$ and the $3N$
numbers
required to specify the (unconstrained) velocities at that point. In a geodesic
problem, one requires respectively $6N$ and $6N-1$ numbers in the two different
but essentially equivalent formulations. In the conformal theory, we thus have
something very like a monotonic `time', but it does not enter as a difference
$t_2-t_1$ but as the ratio $K_2/K_1$. This result seems to me highly
significant because it shows (as just noted in the footnote) that in the
shape-dynamic description of gravity one can
interpret the local shapes of
space as the true degrees of freedom and $K_2/K_1$ as an independent variable.
As $K_2/K_1$ varies, the shapes interact with each other.
This mirrors the
interaction of particle positions in Newtonian dynamics as time, or, as we saw
earlier, $T/V$ changes. However, the closer analogy in Newtonian dynamics is
with the system's change of the shape as $D_2/D_1$ changes.

The only input data in this form of the shape-dynamic conformal theory are
the initial point and tangent vector in CS. There is no trace of local inertial
frames of reference, local proper time, or local proper distance. In the
standard derivation of general
relativity these are all presupposed in the requirement that
locally spacetime can be approximated in a sufficiently small region by
Minkowski space.\footnote{The 4-metric $g_{\mu\nu}$ has 10
components, of which four correspond to coordinate freedom. If one takes the
view, dictated by general covariance, that all
the remaining six are equally physical, then the entire theory rests on
Minkowski space. One merely
allows it to be bent, as is captured in the `comma goes to semicolon' rule.} In
contrast, in the conformal approach, this entire structure emerges from
specification of a point and direction in conformal superspace. 

One or two points may be made in this connection. First, as the reader can see
in \cite{Anderson2005}, the manner in which the theory selects a distinguished
3-geometry in a theory in which only conformal 3-geometry is presupposed
relies on intimate interplay of the theory's ingredients. These are the local
square root and the two different best matchings: wrt to diffeomorphisms and
conformal transformations. Second, the construction of spacetime in a CMC
foliation is fixed to the minutest detail from input that can in no way be
reduced. Expressed in terms of two infinitesimally differing conformal
3-geometries $C_1$ and $C_2$, the outcome of the best
matchings fixes the local scale factor $\sqrt g$ on $C_1$ and $C_2$, making
them into 3-geometries $G_1$ and $G_2$. Thus, it takes one to a definite
position in the conformal orbits. This is the big difference from the best
matching with respect to diffeomorphisms alone and what happens in the particle
model and gauge theory. In these cases the postion in the orbit is not fixed.
Next, the best matching procedure pairs each point on $G_1$ with a unique point
on $G_2$ and determines a duration between them. In the spacetime that the
theory `constructs' the paired points are connected by spacetime vectors
orthogonal to $G_1$ and $G_2$ and with lengths equal to definite
(position-dependent) proper times. These are determined on the basis of the
expression (\ref{cmg}) for the canonical momenta, in which $W=gR$. The lapse
$N$ is $N=\sqrt{T/4R}$ and the amount of proper time $\delta\tau$ between the
paired points is $\delta\tau=N\delta\lambda$. It is obvious that $\delta\tau$
is the outcome of a huge holistic process: the two best matchings together
determine not only which points are to be paired but also the values at the
paired points of all the quantities that occur in the expression
$N=\sqrt{T/4R}$.

We can now see that there are two very different ways of interpreting general
relativity. In the standard picture, spacetime is assumed from the beginning
and it must locally have precisely the structure of Minkowski space. From the 
structural point of view, this is almost identical to an amalgam of Newton's
absolute space and time. This near identity is reflected in the essential
identity locally of Newton's first law and Einstein's geodesic law for the
motion
of an idealized point particle. In both cases, it must move in a straight line
at a uniform speed. As I already mentioned, this very rigid initial structure is
barely changed by Einstein's theory in its standard form. In Wheeler's aphorism
\cite{Misner1973}, ``Space tells matter how to move, matter tells space how to
bend.'' But what we find at the heart of this picture is Newton's first law
barely changed. No explanation for the law of inertia is given: it is a --
one is tempted to say \emph{the} -- first principle of the theory. The
wonderful structure of Einstein's theory as he constructed it rests upon it as a
pedestal. I hope that the reader will at least see that there is another way of
looking at the law of inertia: it is not the point of departure but the
destination reached after a journey that takes into account all possible ways in
which the configuration of the universe could change. 

This bears on the debate about reductionism vs holism. I believe that the
standard spacetime representation of general relativity helps to maintain the
plausibility of a reductionist approach. Because Minkowski's spacetime seems to
be left essentially intact in local regions, I think many people (including
those working in quantum field theory in external spacetimes) unconsciously
assume that the effect of the rest of the universe can be ignored. Well, for
some things it largely can. However, I feel strongly that the creation of
quantum gravity will force us to grasp the nettle. What happens locally is the
outcome of everything in the universe. We already have a strong hint of this
from the classical theory, which shows that the `reassuring' local Minkowskian
framework is determined -- through elliptic equations in fact -- by every last
structural detail in the remotest part of the universe.

\subsection{Shape dynamics or general relativity?}

There is no question that general relativity has been a wonderful success and
as yet has passed every experimental test. The fact that it predicts
singularities is not so much a failure of the theory as an indication that
quantum gravity must at some stage come into play and `take over'. A more
serious criticism often made of general
relativity is that its field equations
$G_{\mu\nu}=T_{\mu\nu}$ allow innumerable solutions that strike one as
manifestly unphysical, for example, the ones containing closed timelike curves.
There is a good case for seeking a way to limit the number of solutions. Perhaps
the least controversial is the route chosen by Dirac \cite{Dirac1958} and
Arnowittt, Deser, and
Misner (ADM) \cite{Arnowitt1962}. The main justification for their 3+1 dynamical
approach is the
assumption that gravity can be described in the Hamiltonian framework, which is
known to be extremely effective in other branches of physics and especially in
quantum mechanics. 

If a Hamiltonian framework is adopted, it then becomes especially attractive to
assume that the universe is spatially closed. This obviates the need for
arbitrary boundary conditions, and, as Einstein put it when discussing Mach's
principle (\cite{Einstein1922}, p. 62), ``the series of causes
of mechanical phenomena [is] closed''. 

The main difficulty in suggesting that the spacetime picture
should be replaced by the more restrictive Hamiltonian framework arises from
the relativity principle, i.e., the denial of any
distinguished definition of simultaneity. In the ideal form of Hamiltonian
theory, one seeks to have the dynamics represented by a \emph{unique} curve in
a phase space of \emph{true Hamiltonian degrees of freedom}. This is equivalent
to having a unique curve in a corresponding configuration space of true
geometrical degrees of freedom even if 
mathematical tractability means that the calculations must always be made in
Riem. Dirac and ADM showed that dynamics in Riem could be interpreted in
superspace, thereby reducing the six degrees of freedom per space point in a
3-metric to the three in a 3-geometry. But the slicing freedom within spacetime
means that a single spacetime still corresponds to an infinity of curves in
superspace. The Hamiltonian ideal is not achieved. The failure is tantalizing,
because much evidence suggests that gravity has only \emph{two} degrees of
freedom per space point, hinting at a configuration space smaller than
superspace.

As long as relativity of
simultaneity is held to be sacrosanct, there is no way forward to the
Hamiltonian ideal. York and Wheeler came close to suggesting that it was to be
found in conformal superspace, but ultimately balked at jettisoning the
relativity principle.\footnote{York's highly important work on the
initial-value problem of general
relativity \cite{York1971, York1972}  is intimately related to
the shape-dynamic
programme and was one of its inspirations. For a discussion of the connections,
see \cite{Anderson2005}. One of the arguments for the shape-dynamic approach is
that it
provides a first-principles \emph{derivation} of York's method, which in its
original form was found by trial and error. It may also be noted here that
York's methods, which were initially developed for the vacuum (matter-free)
case, can be extended to include matter \cite{Isenberg1976, Isenberg1977}. This
suggests that the principles of shape dynamics will extend to the case in which
matter is present.} In this connection, it is worth
pointing out that Einstein's route to general
relativity occurred at a particular point in
history and things could have been approached differently. I think it entirely
possible that Einstein's discovery of his theory of gravitation in spacetime
form could be seen as a glorious historical accident. In particular, Einstein
could easily have looked differently at certain fundamental issues related to
the
nature of space, time, and motion. Let me end this introduction to shape
dynamics with some related observations on each.

\emph{Space}. Riemann based his generalization of Euclidean geometry on
\emph{length} as fundamental. It was only in 1918, three years after the
creation of general
relativity, that Weyl \cite{Weyl1918, Weyl1997} challenged this and
identified -- in a
four-dimensional context -- \emph{angles} as more fundamental. I will argue
elsewhere that Weyl's attempt to generalize general
relativity to eliminate the correctly
perceived weakness of Riemann's foundations failed because it was not
sufficiently radical -- instead of eliminating length completely from the
foundations, Weyl retained it in a less questionable form.

\emph{Time}. As I noted earlier, in 1898 Poincar\'e
\cite{Poincar'e1898, Poincar'e1904} identified \emph{two} equally fundamental
problems related to time: how is one to define \emph{duration} and how is one to
define \emph{simultaneity} at spatially separated points? Einstein attacked the
second problem brilliantly but made no attempt to put a solution to
the second into the foundations of general relativity.

\emph{Motion}. In the critique of Newtonian mechanics that was such a stimulus
to general
relativity, Mach argued that only relative velocities should occur
in dynamics. Einstein accepted this aspiration, but did not attempt to put it
directly into the foundations of general
relativity, arguing that it was impractical
(\cite{Einstein1918, Barbour1995}, p. 186). Instead
it was necessary to use coordinate systems and achieve Mach's ideal by putting
them all on an equal footing (general covariance).

All three alternatives in approach listed above are put directly into shape
dynamics. I think that this has been made adequately clear with regard to the
treatment of space and motion. I
wish to conclude with a comment on the treatment of time, which is rather more
subtle.

It is well known that Einstein regarded special relativity as a principles
theory like
thermodynamics, which was based on human experience: 
heat energy never flows spontaneously from a cold to a hot body. Similarly,
uniform motion was always found to be indistinguishable --
within a closed system -- from rest. Einstein  took this fact as the basis
of relativity and never attempted to explain
effects like time dilatation at a microscopic level in the way Maxwell and
Boltzmann developed the atomic statistical theory of thermodynamics. Since rods
and clocks are ultimately quantum objects, I do not think such a programme can
be attempted before we have a better idea of the basic structure of quantum
gravity. However, I find it interesting and encouraging that \emph{a microscopic
theory of duration} is built in at a very basic level in shape dynamics. This
is achieved in particle dynamics using Jacobi's principle, which leads to a
global definition of duration, and in conformal dynamics using the
local-square-root action (\ref{sqrt1}). I also find it striking that, as
already noted, the simple device of eliminating the lapse from Einstein's
spacetime theory immediately transforms his theory from one created without any
thought of a microscopic theory of duration into one (based on the
Baierlein--Sharp--Wheeler action (\ref{BSW})) that has such a theory at its
heart. A theory of duration was there all along. It merely had to be uncovered
by removing some of the structure that Einstein originally employed -- truly a
case of less is more.

The effect of the local square root is remarkable. At the level of theory
creation in superspace, in which length is taken as fundamental, the local
square root acts
as an extremely powerful selector of consistent theories and, as we have seen,
enforces the appearance of the slicing freedom, universality of the light cone,
and gauge fields as the simplest bosonic fields that couple to dynamic
geometry. As I have just noted, it also leads to a microscopic theory of local
duration (local proper
time). Thus, the mere inclusion of the local square root goes a long way to
establishing a
constructive theory of special-relativistic effects. It is not the whole
way, because quantum mechanics must ultimately explain why physically realized
clocks measure the local proper time created by the local theory of duration.

The effect of the local square root is even more striking when applied in theory
creation in
conformal superspace. It still enforces universality of the light cone and
the appearance of gauge fields but now does two further things. First, it leads
to a \emph{microscopic theory of length}. For the conformal best matching, in
conjunction with the constraints that follow from the local square root, fixes a
distinguished
scale factor of the 3-metric. Second, it introduces the distinguished
CMC foliation within spacetime without changing any of the classical
predictions of general relativity. It leads to a \emph{theory of simultaneity}.

Thus, the conformal approach to geometrodynamics suggests that there are two
candidate theories of gravity that can be derived from different first
principles. Einstein's general relativity is based on the idea that spacetime is
the basic ontology; its symmetry group is four-dimensional diffeomorphism
invariance. But there is also an alternative \emph{dual} theory based
on
three-dimensional diffeomorphism invariance and conformal best
matching \cite{Gomes2011, Gomes2011a, Anderson2005, Barbour2010a}. The set
of
allowed solutions of the conformal theory is significantly smaller than the
general
relativity
set. In principle, this is a good feature, since it makes the conformal
theory more predictive, but it cannot be ruled out that, being tied to CMC
foliations, the conformal theory will be unable to describe physically
observable situations that are correctly described by general
relativity.

I will end with two comments. First,
shape dynamics in conformal superspace is a new and mathematically well-defined
framework of dynamics. Second, its physical applications are most likely to
be in quantum gravity.
\vspace{.2in} 

\bf{Acknowledgements.} The work reported in this paper grew out of
collaboration with (in chronological order) Bruno Bertotti, Niall \'O Murchadha,
Brendan Foster, Edward Anderson, Bryan Kelleher, Henrique Gomes, Sean Gryb, and
Tim Koslowski. Discussions over many years with Karel Kucha\v r and Lee Smolin
were also very helpful, as were extended discussions with Jimmy York in 1992. I
am much indebted to them all. Thanks also to Boris Barbour for creating the
figures. This work was funded by
a grant from the Foundational
Questions Institute (FQXi) Fund, a donor advised fund of the Silicon Valley
Community Foundation on the basis of proposal FQXi-RFP2-08-05 to the
Foundational Questions Institute.

\bibliography{/home/julian/latex/bibtex/julian_barbour_references}
\end{document}